\documentclass[12pt]{article}

\usepackage{color}
\usepackage[dvips]{graphicx,psfrag}
\usepackage{amsmath,amssymb}
\usepackage{bm}

\baselineskip=\normalbaselineskip
\renewcommand{\baselinestretch}{1.34}
\renewcommand{\thefootnote}{\fnsymbol{footnote}}

\newcommand{\vev}[1]{\left\langle #1 \right\rangle}
\newcommand{\ket}[1]{\bigl|#1\bigr>}
\newcommand{\bra}[1]{\bigl<#1\bigr|}

\newcommand{\maru}[1]
{{\ooalign{\hfil#1\/\hfil\crcr\raise.167ex\hbox{\mathhexbox20D}}}}

\newcommand{\cH}{\mathcal{H}}

\newcommand{\cL}{\mathcal{L}}
\newcommand{\cM}{\mathcal{M}}

\newcommand{\cO}{\mathcal{O}}
\newcommand{\cP}{\mathcal{P}}
\newcommand{\cQ}{\mathcal{Q}}

\newcommand{\cS}{\mathcal{S}}

\newcommand{\cV}{\mathcal{V}}
\newcommand{\cW}{\mathcal{W}}

\newcommand{\cZ}{\mathcal{Z}}

\newcommand{\bC}{\mathbb{C}}

\newcommand{\bR}{\mathbb{R}}
\newcommand{\bZ}{\mathbb{Z}}

\newcommand{\bP}{\boldsymbol{P}}
\newcommand{\bQ}{\boldsymbol{Q}}
\newcommand{\bPone}{\boldsymbol{P_1}}
\newcommand{\bPtwo}{\boldsymbol{P_2}}
\newcommand{\bQone}{\boldsymbol{Q_1}}
\newcommand{\bQtwo}{\boldsymbol{Q_2}}

\newcommand{\bunit}{\boldsymbol{1}}
\newcommand{\bH}{\boldsymbol{H}}
\newcommand{\bA}{\boldsymbol{A}}

\newcommand{\bL}{\boldsymbol{L}}

\newcommand{\bW}{\boldsymbol{W}}

\newcommand{\be}{\boldsymbol{e}}

\newcommand{\bsigma}{\boldsymbol{\sigma}}

\newcommand{\bPhi}{\boldsymbol{\Phi}}
\newcommand{\del}{\partial}

\newcommand{\eq}[1]{(\ref{#1})}

\newcommand{\nn}{\nonumber}

\DeclareMathOperator{\tr}{tr}

\allowdisplaybreaks[1]

\newcommand{\gint}{\oint\hspace{-14pt}\bigcirc}

\newcommand{\ds}{\displaystyle}
\newcommand{\nol}
  {\!
    \begin{array}{r}
    \raisebox{-1.7mm}{\mbox{\scriptsize{$\circ$}}} \\ 
    \raisebox{2.7mm}{\mbox{\scriptsize{$\circ$}}}
    \end{array}
   \!\!}
\newcommand{\nor}
  {\!\!
    \begin{array}{l}
    \raisebox{-1.7mm}{\mbox{\scriptsize{$\circ$}}} \\ 
    \raisebox{2.7mm}{\mbox{\scriptsize{$\circ$}}}
    \end{array}
   \!}
   
\newcommand{\uP}{U^{P}}
\newcommand{\uQ}{U^{Q}}

\makeatletter
\@addtoreset{equation}{section}
\@addtoreset{theorem}{section}
\@addtoreset{definition}{section}
\@addtoreset{lemma}{section}
\@addtoreset{proposition}{section}

\begin{document}
\begin{flushright}
\parbox{40mm}{%
KUNS-2047 \\
{\tt hep-th/0611045} \\
November 2006}
\end{flushright}

\vfill

\begin{center}
{\Large{\bf 
A string field theoretical description \\
of $(p,q)$ minimal superstrings
}}
\end{center}

\vfill

\begin{center}
{\large{Masafumi Fukuma}}\footnote%
{E-mail: {\tt fukuma@gauge.scphys.kyoto-u.ac.jp}} and  \!
{\large{Hirotaka Irie}}\footnote%
{E-mail: {\tt irie@gauge.scphys.kyoto-u.ac.jp}}   \\[2em]
Department of Physics, Kyoto University, 
Kyoto 606-8502, Japan \\

\end{center}
\vfill
\renewcommand{\thefootnote}{\arabic{footnote}}
\setcounter{footnote}{0}
\addtocounter{page}{1}

\begin{center}
{\bf abstract}
\end{center}

\begin{quote}
A string field theory of $(p,q)$ minimal superstrings 
is constructed with the free-fermion realization of 
2-component KP (2cKP) hierarchy, 
starting from 2-cut ansatz of two-matrix models. 
Differential operators of 2cKP hierarchy are identified 
with operators in super Liouville theory, 
and we obtain algebraic curves for the disk amplitudes of $\eta=-1$ FZZT-branes  
and the partition functions of neutral/charged $\eta=-1$ ZZ branes, 
which correctly reproduce those of type 0B $(p,q)$ minimal superstrings 
in conformal backgrounds. 
In the course of study, 
some subtle points are clarified, 
including a difference of $(p,q)$ even/odd models 
and quantization of flux, 
and we show that the Virasoro constraints naturally 
incorporate quantized fluxes without ambiguity. 
We also argue within this string field framework 
that type 0A minimal superstrings can be obtained 
by orbifolding the type 0B strings with a $\bZ_2$ symmetry 
existing when special backgrounds are taken. 
\end{quote}
\vfill
\renewcommand{\baselinestretch}{1.4}
\newpage

\section{Introduction}

Noncritical string theories \cite{Polyakov} are useful toy models 
to investigate various aspects of superstrings. 
While sharing many of important properties 
with their critical counterparts,  
noncritical string theories have fewer degrees of freedom 
and in many cases they are integrable \cite{KPZ,DHK,dsl}. 
They also have a description based on a string field theory 
for bosonic cases \cite{fy1,fy2,fy3,FIS,fim}.

For the last ten years, a great progress has been made 
in the understanding of noncritical superstring theories 
based on super Liouville theory 
\cite{DOZZ,fzz-t,zz,superLiouville, Mar, SeSh,Oku}, 
and some exact results on correlation functions have been obtained. 
The main aim of the present paper 
is to construct a string field theory 
of $(p,q)$ minimal superstrings 
such that it has a nonperturbative definition  
and correctly reproduces all the known results 
in super Liouville field theory.

Our basic strategy is to use 2-cut solutions of 
two-matrix models as a nonperturbative definition 
of $(p,q)$ minimal superstring theories. 
This is based on the recent observation 
that 2-cut solutions \cite{GroWit,PeShe,Napp,CDM,HMPN,ogu,BDJT} 
of one-matrix models with symmetric double-well 
potentials describe two-dimensional pure supergravity, 
or $(p,q)=(2,4)$ type 0 superstrings \cite{UniCom,SeSh2}.  
There the symmetric fluctuations of eigenvalues are interpreted 
as being in the NS-NS sector while the antisymmetric ones 
in the R-R sector \cite{tt,newhat}.
Furthermore, by fine-tuning the potentials,  
one obtains a series of higher multicritical points, 
which are identified with $(p,q)=(2,4k)$ type 0 superstrings \cite{UniCom}.%
\footnote{There are also $(2,4k+2)$ theories in a flow generated by R-R operators.} 
This is reminiscent of what happened for bosonic minimal string theory, 
where one-cut solutions of one-matrix models describe 
$(p,q)=(2,2k-1)$ Kazakov series \cite{Kazakov:1989bc,Staudacher}, 
while those of two-matrix models describe all the $(p,q)$ minimal strings 
\cite{Douglas:1989dd, 2-mat}. 
Thus, it is natural to expect that 
generic $(p,q)$ minimal superstring theories can be defined 
as continuum limits of two-matrix models in 2-cut phases. 
We show that this is indeed the case; 
we find that 2-cut solutions of two-matrix models 
generically have the integrable structure of 
the 2-component KP (2cKP) hierarchy, 
and that physical operators in super Liouville theory 
have their counterparts in the 2cKP hierarchy. 
We will see that this mapping enables us to construct 
a string field theory of $(p,q)$ minimal superstrings, 
and the obtained results correctly reproduce those
in super Liouville theory.

We here give an intuitive explanation of 
why 2-cut solutions of matrix models 
and the 2cKP hierarchy play a particular role 
in type 0 superstring theory. 

Matrix models are originally introduced 
as the generators of random triangulations of a surface 
and are found to successfully describe bosonic noncritical strings 
(or two-dimensional quantum gravity) 
in their continuum limits.
The matrix models of type 0 superstrings do not give 
triangulations of a surface in a usual sense 
but behave as the generators of ``the square roots of triangulations." 
This can be seen as follows.

First, from the worldsheet side 
with Liouville superfield 
$\Phi(z,\bar z,\theta,\bar\theta)
=\phi(z,\bar z)+i\theta\psi(z,\bar z)
 +i\bar\theta\bar\psi(z,\bar z)
+i\theta\bar\theta F(z,\bar z)$, 
the bulk cosmological constants $\mu$ is introduced 
as the coefficients of super Liouville potential (see, e.g., \cite{newhat}): 
\begin{align}
 S_{\rm int}^{\rm (bulk)}
  &\sim \int dz d\bar z d\theta d\bar\theta\,\bigl[ -i\mu\,
  e^{ b\Phi(z,\bar z,\theta,\bar\theta)} \bigr]
  \sim \int dz d\bar z\,\bigl[
   -i\mu\,\psi \bar\psi\,
   e^{b\phi}+(\mu^2/4)\,e^{2\,b\,\phi}\bigr],
\end{align} 
from which one sees that $\mu$ is the square root 
of its bosonic string counterpart, $\mu_{ bos}$, 
the Laplace conjugate to the area 
of random surface. 
A similar analysis can be made also for the boundary cosmological 
constant $\zeta$ and its bosonic counterpart $\zeta_{ bos}$, 
and we have the relation
\begin{align}
 \zeta\sim \sqrt{\zeta_{\it bos}}, \qquad \mu \sim \sqrt{\mu_{\it bos}}.
 \label{double1}
\end{align}
That is, the boundary and bulk cosmological constants of minimal superstrings 
are related to the square roots of their bosonic counterparts. 

On the other hand, the boundary cosmological constant $\zeta_{ bos}$ 
can be regarded as the complex coordinate 
of two-dimensional target space \cite{fy3}. 
In fact, in the bosonic case,  
the macroscopic operator 
$\ds\cO(\zeta_{ bos})=\int_0^\infty\!dl\,e^{-l\,\zeta_{\it bos}}\,\cO(l)$ 
can be expressed 
(up to the so-called nonuniversal terms) 
as the summation of microscopic operators 
$\cO_n$ 
\begin{align}
 \cO(\zeta_{\it bos})
  \sim\sum_{n}\,c_n\,\cO_n\,\zeta_{\it bos}^{-n/p-1},
 \label{O_zeta}
\end{align}
where $\ds\cO_n=\int \!dzd\bar z \,
e^{\,i\alpha_n X(z,\bar z)+\beta_n\,\phi(z,\bar z)}$
with the Feigin-Fuchs matter $X(z,\bar z)$, 
and $\alpha_n=(b^{-1}-b)-n/\sqrt{pq}$ 
and $\beta_n=(b^{-1}+b)-n/\sqrt{pq}$ 
for $c_{\rm matter}=1-6(q-p)^2/qp$ and $b=\sqrt{p/q}\,(<1)$. 
The coefficients $c_n$ may be complicated, 
but in any case $\cO(\zeta_{bos})$ receives a nonnegligible contribution 
when the following relation holds for some $(z,\bar z)$:
\begin{align}
 \zeta_{bos}^{1/p}\sim e^{-b\,
  \bigl(\phi(z,\bar z)+i\,X(z,\bar z)\bigr)}. 
 \label{double2}
\end{align} 
This enables us to interpret the operator $\cO(\zeta_{bos})$ 
as creating a closed loop at the spacetime complex coordinate 
$\zeta_{bos}$. 
This also implies that $\zeta_{bos} \in \mathbb R_+$ is related to 
Liouville coordinate $\phi$ and the limit ${\rm Re}\,\zeta_{bos}\to+\infty$ 
corresponds to the weak coupling region $\phi\to-\infty$ \cite{MMSS}.%
\footnote{
We should stress here that the above is a rough discussion 
since terms with large $n$ in the summation \eq{O_zeta} become nonnegligible  
when $|\zeta_{bos}|$ is small enough which may be out of the perturbative region. 
}

In super Liouville theory, it is again $\zeta_{bos}\sim\zeta^2$ 
that allows such analysis based on the Feigin-Fuchs representation. 
Thus, for a given spacetime Liouville coordinate $\zeta_{bos}\in \mathbb R_+$, 
the boundary cosmological constants $\zeta$ and $-\zeta$ are naturally paired 
to give the bosonic boundary cosmological constant $\zeta_{bos}\sim\zeta^2$, 
and the weak coupling region of Liouville theory, $\phi\to -\infty$, 
corresponds to ${\rm Re}\, \zeta \to \pm \infty$. 

Since the cuts in a spacetime geometry of $\zeta$ are also paired to give 
a single cut for bosonic counterpart $\zeta^2\sim\zeta_{bos}$ (Fig.\ 1), 
such a geometry is realized in a 2-cut hermitian matrix model with 
a symmetric double-well potential, 
such that its eigenvalue $x\in\bR$ in one cut 
is always paired with $-x$ in the other cut. 
It then can be understood that $\zeta$ in super Liouville theory 
is related to the matrix-model eigenvalue $x$ as
\begin{align}
 \zeta=-i x, 
\end{align}
which was first pointed out in \cite{UniCom}.

\begin{figure}[htbp]
\begin{center}
\resizebox{!}{45mm}{
\input{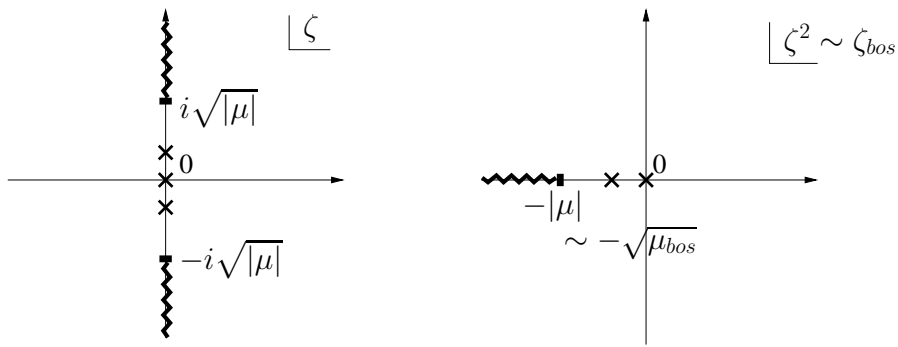}}
\end{center}
\caption{\footnotesize{
A typical geometry of spacetime with $\zeta$ and $\zeta^2$
}}
\label{2cut}
\end{figure}%

From the viewpoint of integrable system  
(or string field theory discussed in this paper), 
the coordinate $\zeta_{\it bos}$ in bosonic string theories 
is represented as the eigenvalue of a differential operator 
$\bP=\del^p+\cdots$ (see, e.g., \cite{fim}). 
Such a system including higher-order differential operators 
can be systematically analyzed 
once it is embedded into the KP hierarchy 
(see, e.g., \cite{sato-sato,djkm}).
For type 0 NSR superstrings, 
we consider 
a $2\times 2$ matrix-valued differential operator of the form
\begin{align}
 \bP =\sigma_3\,\del^{\hat p} + \cdots
\end{align}
which naturally has a pair of eigenfunctions $\Psi^{(1)}$ and $\Psi^{(2)}$ 
with eigenvalues $\zeta$ and $-\zeta$, respectively:
\begin{align}
 \bP\,\Psi^{(1)} = \zeta\,\Psi^{(1)},\qquad 
 \bP\,\Psi^{(2)} = -\zeta\,\Psi^{(2)}.
\end{align}
This is reminiscent of the Dirac theory of fermions, 
where the ``square root'' of the Klein-Gordon equation is realized 
by using a matrix-valued differential operator. 
Such a system turns out to be in a class of 2cKP hierarchy 
as we review in the next section. 
There we will find that the operators 
of the form $\boldsymbol{O}^{[0]}_n=\del^n+\cdots$ 
correspond to the fluctuations of an NS-NS scalar, 
and those of the form $\boldsymbol{O}^{[1]}_n=\sigma_3\,\del^n+\cdots$ 
to the fluctuations of an R-R scalar. 
We will also see that such system can be fully described 
by 2-component fermions,
\begin{align}
 \bigl(c_0^{(1)}(\zeta),\bar c_0^{(1)}(\zeta)\bigr), \quad 
 \bigl(c_0^{(2)}(\zeta),\bar c_0^{(2)}(\zeta)\bigr), 
\end{align}
the former of which corresponds to the creation/annihilation of 
an $\eta=-1$ FZZT brane of charge $+1/2$  
and the latter to that of charge $-1/2$, 
both placed at the same spacetime point $\zeta^2$ 
but at different ``superspace points'' $\zeta$ and  $-\zeta$, respectively.%
\footnote{
Here $\eta$ represents the relation of the left and right supercharges 
on a boundary, 
$Q_{\rm L}=i\eta\,Q_{\rm R}$. 
The operators that can be naturally derived from matrix models 
are found to choose $\eta=-1$, as was pointed out in \cite{UniCom}. 
It would be interesting to investigate 
whether our string field theory can describe the case $\eta=+1$. 
There is an interesting proposal from a loop gas approach \cite{GRT}
} 

This paper is organized as follows. 
In section 2, we analyze two-cut two-matrix models 
closely following the analysis made in one-cut two-matrix models 
\cite{Douglas:1989dd,2-mat,fkn3}, 
and show that operators in super Liouville field theory 
can totally be written in terms of 2cKP hierarchy. 
In section 3, we solve the Douglas equation in such systems 
and derive the $W_{1+\infty}$ constraints. 
We also derive string equations, 
and show that background R-R fluxes can be incorporated 
into our formalism 
without ambiguity (not as integration constants in string equations). 
This reflects the fact that the Virasoro constraints 
(included in the $W_{1+\infty}$ constraints) 
are in the ``once-integrated form'' of string equations \cite{fkn1}. 
In section 4, we introduce type 0B $(p,q)$ minimal string field theory, 
and clarify the meaning of FZZT and ZZ branes in our context. 
In section 5, we calculate the disk amplitudes of 
$\eta=-1$ FZZT branes and derive the corresponding algebraic curves. 
We find there that the even and odd minimal superstrings, 
though apparently different, 
can have a common description. 
We also calculate the partition functions of neutral/charged $\eta=-1$ 
ZZ branes. Section 6 is devoted to conclusion and discussion, 
and we there make a comment on how type 0A superstrings 
(which is described by complex matrix models \cite{DJM,uni2dqg,LM,DiF}) 
are obtained within our framework.

\section{2-cut two-matrix models and 2cKP hierarchy }

In this section, we investigate 2-cut two-matrix models and show that 
2-cut ansatz gives 2-component KP (2cKP) hierarchy 
\cite{sato-sato,djkm2c,KaLe}  as their integrable structure. 
We also establish the identification of the differential operators 
appearing in 2cKP hierarchy with the operators 
in $(p,q)$ minimal superstring theories.

\subsection{Two-matrix models with 2-cut ansatz}\label{2-cut_ansatz}

The partition function of a two-matrix model is 
defined by the following integral of two $N\times N$ 
hermitian matrices $X$ and $Y$:
\begin{align}
 Z_{\rm lat}\equiv \int dX dY e^{-N\tr w(X,Y)},
  \quad w(X,Y)\equiv V_1(X)+V_2(Y)-cXY,
\end{align} 
and can be written in terms of the eigenvalues of $X$ and $Y$ 
($\{x_i\}$ and $\{y_i\}$ ($i=1,\cdots,N$), respectively) 
as 
\begin{align}
 Z_{\rm lat}&=\int\! \prod_{i=1}^N dx_i\,dy_i\,
  \Delta(x)\,\Delta(y)\,
  e^{-N \sum_i w(x_i,y_i)}. 
\end{align}
Here $\Delta(x)$ and $\Delta(y)$ are 
the van der Monde determinants 
$\bigl({\rm e.g.\,}\ds\Delta(x)=\prod_{i<j}(x_i-x_j)\bigr)$. 
Since 2-cut critical points of one-matrix models 
were found at the origin of $\bZ_2$ symmetric potential 
of eigenvalue $\lambda$, $V(-\lambda)=V(\lambda)$ \cite{GroWit,PeShe,Napp}, 
it is natural to expect that 2-cut critical points of two-matrix models 
are obtained from the following $\mathbb Z_2$ symmetric potential%
\footnote{
In a similar way, we can introduce multi-cut, multi-critical multi-matrix models 
with the strategy of \cite{multcmultc}. 
For example, 4-cut two-matrix models can be derived by requiring 
the following $\mathbb Z_4$ symmetry: 
$w(\omega x,\omega^3 y)=w(x,y) , \, (\omega=e^{2\pi i/4})$, 
under which the interaction term $cxy$ is invariant.
}%
\begin{align}
w(-x,-y)=w(x,y), \label{Z2sym} 
\end{align}
or equivalently $V_1(-x)=V_1(x)$ and $V_2(-y)=V_2(y)$. 
In this paper, we consider more general cases where systems may also be under 
$\bZ_2$-odd perturbations, and study the two-matrix potential 
with the following $\mathbb Z_2$ property:
\begin{align}
w(-x,-y)=w^*(x,y), \label{Z2str}
\end{align}
where ${}^*$ means complex conjugation.
This complex conjugation implies that coefficients of 
odd terms in $V_1(x)$ and $V_2(y)$ are pure imaginary. 
This follows from the one-matrix-model analysis \cite{HMPN}, 
where critical points are found on flows of $\bZ_2$-odd perturbations 
only when the coefficients of odd terms are pure imaginary. 
It is an analogue of Lee-Yang edge singularity \cite{Staudacher}, 
and matrix models with this setup are known to be defined nonperturbatively 
with potentials bounded from below. 
We assume that it is also the case in two-matrix models.

This model can be solved by investigating 
its orthogonal polynomial system 
\begin{align}
 \alpha_n(x)=\frac{1}{\sqrt{h_n}}\,\bigl(x^n+\cdots\bigr),\quad
  \beta_n(x)=\frac{1}{\sqrt{h_n}}\,\bigl(y^n+\cdots\bigr)
  \qquad (n=0,1,2,\cdots),
\end{align}
which we now require to satisfy the orthonormality conditions:
\begin{align}
 \delta_{m,n}=\bigl<\alpha_m\bigr|\beta_n\bigr>
  \equiv \int\!dx\,dy\,e^{-N w(x,y)}\,\alpha_m(x)\,\beta_n(y),
\end{align}
and the partition function is given as 
$ Z_{\rm lat}= N!\,\prod_{n=0}^{N-1}\,h_n $.
Reflecting the above $\bZ_2$ symmetry \eq{Z2str}, 
the orthonormal polynomials $\alpha_n(x)$ and $\beta_n(y)$
are equipped with the following $\mathbb Z_2$ structure:
\begin{align}
 \alpha_n(-x)=(-1)^n\,\alpha_n^*(x),\quad
  \beta_n(-y)=(-1)^n\,\beta_n^*(y)
  \qquad (n=0,1,2,\cdots).
\end{align}
Then the 2-cut critical points are realized 
at the origin of eigenvalues, $(x_c,y_c)=(0,0)$.%
\footnote{
More generally, one can assume $(x_c,y_c)=(ic_1,ic_2)$, 
but this always can be shifted to the origin 
by a proper deformation of potential, 
which corresponds to redundant operators in the R-R sector. 
} 

This polynomial system is characterized 
by the canonical pairs of operators 
$\bQone$, $\bPone$, $\bQtwo$ and $\bPtwo$ defined by 
\begin{align}
 x\,\alpha_n(x)
  &=\sum_m\,\alpha_m(x)\,\bigl(\bQone\bigr)_{mn},\qquad
 \frac{d}{dx}\alpha_n(x)
  =\sum_m\,\alpha_m(x)\,\bigl(\bPone\bigr)_{mn},\\
 y\,\beta_n(y)
  &=\sum_m\,\beta_m(y)\,\bigl(\bQtwo\bigr)_{mn},\qquad
 \frac{d}{dy}\beta_n(y)
  =\sum_m\,\beta_m(y)\,\bigl(\bPtwo\bigr)_{mn},
\end{align}
which satisfy the relations
\begin{align}
 \bigl[\bPone,\,\bQone\bigr]=\bunit,\qquad
  \bigl[\bPtwo,\,\bQtwo\bigr]=\bunit.
\end{align}
They are equivalent to
\begin{align}
 \bigl[\bQone^{\rm T},-c\bQtwo\bigr]=N^{-1}\,\bunit,
 \label{douglas_lat}
\end{align}
since $\bPone$ and $\bPtwo$ satisfy
\begin{align}
 \bPone=N\bigl(-c\,\bQtwo^{\rm T}+V_1'(\bQone)\bigr),\quad 
 \bPtwo=N\bigl(-c\,\bQone^{\rm T}+V_2'(\bQtwo)\bigr). \label{EOMofPQ}
\end{align}
Since the scaling behaviours of these polynomials 
are different for even and odd index $n$, 
we rearrange the orthonormal basis such that $n$ runs 
as $n=0,2,4,\cdots,1,3,5,\cdots$ and multiply the basis $(\alpha_{2k}(x),$
$\alpha_{2k+1}(x))$ and $(\beta_{2k}(y),$ $\beta_{2k+1}(y))$ 
by a common factor $(-1)^k$ \cite{multcmultc}. 
Then the operators $\bQ_1^{\rm T}$ and $\bQ_2$ 
become 2$\times$2 block matrices, each of four elements is 
a difference operator, and we can render these difference operators 
such as to behave smoothly for a small change of $n$ by tuning the 
potentials properly.

As is discussed in Appendix \ref{Douglas} in detail 
we can take a continuum limit in such a way that 
they have the following scaling behaviors 
with respect to the lattice spacing $a$ of random surfaces:
\begin{align}
 N^{-1}=g\,a^{(\hat{p}+\hat{q})/2},\qquad
 \bQone=
  ic_1\bunit_2 + i\,a^{\hat{p}/2}\,\bP,\qquad
 \bQtwo^{\rm T}=
  ic_2\bunit_2 + i\,a^{\hat{q}/2}\,\bQ. 
 \label{PandQ}
\end{align}
Here $\bP$ and $\bQ$ are $2\times 2$ matrix-valued differential operators 
of order $\hat{p}$ and $\hat{q}$, respectively, 
with respect to the scaling variable 
$\xi\equiv -a^{-(\hat{p}+\hat{q}-1)/2}\dfrac{N-n}{N}$ 
$(<0)$: 
\begin{align}
 \bP=\sum_{i=0}^{\hat{p}}\,\uP_i(\xi)\,\del^{\hat{p}-i},\quad
  \bQ=\sum_{j=0}^{\hat{q}}\,\uQ_j(\xi)\,\del^{\hat{q}-j}\qquad
  \Bigl(\del\equiv g\,\frac{\del}{\del \xi}
   = -a^{-1/2}\frac{\del}{\del n}\Bigr).
 \label{PQdel}
\end{align}
The analysis made in Appendix \ref{Douglas} 
allows us to set the following reality condition 
on the coefficients%
\begin{align}
\uP_k, \ \uQ_k\in \mathbb R \,\bunit_2+\mathbb R \,\sigma_3 
+ \mathbb R \,\sigma_2 +i\,\mathbb R \,\sigma_1, \label{OpHer}
\end{align}
especially $\uP_0,\, \uQ_0 \in \mathbb R\, \bunit_2+\mathbb R\, \sigma_2$. 
Then eq.\ \eq{douglas_lat} is rewritten into the form 
of the Douglas equation \cite{Douglas:1989dd}
\begin{align}
 \bigl[\bP,\,\bQ\bigr]=g\,\bunit.
 \label{douglas-eq}
\end{align}

The Douglas equation \eq{douglas-eq} is invariant under 
the transformation %
\begin{align}
 \bP\to c\,V(\xi)\cdot\bP\cdot V(\xi)^{-1},\qquad
  \bQ\to c^{-1}\,V(\xi)\cdot\bQ\cdot V(\xi)^{-1}
\end{align}
with a nonvanishing constant $c$ 
and a regular function of $\xi$, 
$V(\xi)=e^{\,\bunit_2 f_1(\xi)+\sigma_3f_2(\xi)}e^{i(\pi/4)\sigma_1}$. 
Using this, we can always assume that 
the pair $(\bP, \bQ)$ is in the canonical form which satisfies:
\begin{align}
 \uP_0=\sigma_3^{\epsilon}e^{\theta_p\sigma_3},
 \quad \uQ_0=c_q\,(\sigma_3\cos\theta_q+\bunit_2\sin\theta_q), 
 \quad \uP_1=
 \begin{pmatrix}
 0&{\uP_+}\\
 {\uP_-}&0
 \end{pmatrix}.
 \label{generic}
\end{align}
The case of $\theta_p=0$ and $\epsilon=1$ is of our concern 
as was discussed in Introduction, and we assume this 
in the rest of this article:
\footnote{
This assumption can actually be set without loss of generality. 
See footnote \ref{generic2} in subsection \ref{Winf_constr}.
}
\begin{align}
 \bP=\sigma_3\,\del^{\hat p}+\cdots,
 \quad \bQ=\bigl(c_q^{[0]}\sigma_3+c_q^{[1]}\bunit_2\bigr)\del^{\hat q}
  +\cdots.
 \label{PQ_general}
\end{align}
It turns out that the special cases where $c_q^{[1]}=0$  
correspond to critical points obtained from $\mathbb Z_2$ symmetric 
potentials (i.e.\ universality classes defined by NS-NS operators), 
and that the cases where $c_q^{[0]}=0$ correspond to critical points 
to be found in odd perturbations 
(i.e.\ in a flow generated by R-R operators).

\subsection{Deformation of potentials and 2cKP hierarchy}

Under deformations of the potential $w(x,y)$ in two-matrix models
\begin{align}
 N\,w(x,y) \to N\,w(x,y)+N\,\delta w(x,y),
\end{align}
the matrix-valued differential operators $\bP$ and $\bQ$ will change as 
\begin{align}
 \delta\bP=\frac{1}{g}\bigl[\bH,\bP\bigr],\qquad
 \delta\bQ=\frac{1}{g}\bigl[\bH,\bQ\bigr],
 \label{deform}
\end{align}
with retaining the Douglas equation \eq{douglas-eq}.
An analysis similar to the 1-cut case \cite{sato-sato, SW} shows
that such $\bH$ can be expanded as%
\footnote{
By requiring that $\bP+\delta\bP$ and $\bQ+\delta\bQ$ also retain 
the $\mathbb Z_2$ structure \eq{Z2str}, 
all the parameter $\delta x_n^{[\mu]}$ must be real, 
and thus $\bH$ also satisfies the condition \eq{OpHer}.
} 
\begin{align}
\bH=\sum_{n=0}^{\infty}\sum_{\mu=0,1}\,\delta x_n^{[\mu]}\, 
\bigl(\bsigma^\mu\bL^n\bigr)_+.
\end{align} 
Here $\bsigma$ and $\bL$ are matrix-valued pseudo-differential operators 
of the form
\begin{align}
\bsigma=\sigma_3+\sum_{n=1}^\infty Z_n(\xi)\,\del^{-n},\quad 
 \bL=\bunit_2\,\del+\sum_{n=1}^\infty U_{n+1}(\xi)\,\del^{-n}
\end{align}
which satisfy the relations%
\footnote{
In this paper, $\bsigma^\mu$ means $(\bsigma)^\mu$, and 
the index of the Pauli matrix is denoted by subscript: 
$
\sigma_1=
\begin{pmatrix}
0 & 1 \cr 1 & 0
\end{pmatrix}
$,
$
\sigma_2=
\begin{pmatrix}
0 & -i \cr i & 0
\end{pmatrix}
$ and
$
\sigma_3=
\begin{pmatrix}
1 & 0 \cr 0 & -1
\end{pmatrix} 
$. 
For example, $\sigma_3^2=(\sigma_3)^2=1$ etc.
} 
\begin{align}
\bigl[\bsigma,\bP\bigr]=0,\quad \bsigma^2=\bunit_2,\quad 
\bsigma \bL^{\hat{p}}=\bP,
 \label{p-reduction}
\end{align}
and the positive and negative parts of a matrix-valued pseudo-differential 
operator 
$ \bA=\sum_{n\in\bZ} a_n \del^n$ 
are defined as
\begin{align}
 \bA_+\equiv \sum_{n\geq0} a_n \del^n,\qquad
  \bA_-\equiv \sum_{n<0} a_n \del^n.
\end{align}
One can easily see that the operators of $(\bL^n)_+$ $(\mu=0)$ 
are given by even deformations of the potential $w(x,y)$ 
while $(\bsigma\bL^n)_+$ $(\mu=1)$ by odd deformations.

By assuming that the pair of differential operators $(\bP,\bQ)$ 
are under perturbations with finite $x^{[\mu]}_n$'s, 
they become functions of the variables 
$x=(x_n^{[\mu]})$ $(n=1,2,\cdots;\,\mu=0,1)$, 
and the deformation equations are rewritten as 
\begin{align}
 g\,\frac{\del\bP}{\del x_n^{[\mu]}}=\bigl[(\bsigma^\mu\bL^n)_+,\bP\bigr],
 \qquad 
 g\,\frac{\del\bQ}{\del x_n^{[\mu]}}=\bigl[(\bsigma^\mu\bL^n)_+,\bQ\bigr].
 \label{deforPQ}
\end{align}
One can easily find that the operators $\bsigma$ and $\bL$ 
are uniquely determined for a given differential operator $\bP$, 
and thus the deformation equations of $\bsigma$ and $\bL$ are given by
\begin{align}
g\,\frac{\del\bsigma}{\del x_n^{[\mu]}}
=\bigl[(\bsigma^\mu\bL^n)_+,\bsigma\bigr],
 \qquad 
 g\,\frac{\del\bL}{\del x_n^{[\mu]}}&=\bigl[(\bsigma^\mu\bL^n){}_+,\,\bL\bigr].
  \label{2cKP}
\end{align}
These are known as the Lax equations of 2cKP hierarchy.
Thus we conclude that the underlying integrable structure 
of 2-cut two-matrix models is the 2cKP hierarchy. 
Note that $\del\,\bigl(\equiv\del/\del\xi\bigr)=g\,\del/\del x_1^{[0]}$ 
since $\bL_+=\bunit_2\,\del$. 
Thus, all of the above operators depend on the indeterminate $\xi$ as 
$f(\xi,x)=f(x_1^{[0]}+g\xi,x_2^{[0]},\cdots; \, x_1^{[1]},x_2^{[1]},\cdots)$. 
Therefore, in the rest of the present paper, 
we absorb $\xi$ into $x^{[0]}_1$ and set $\xi=0$.

The 2cKP hierarchy equations \eq{2cKP} are the conditions 
that ensure the existence of a matrix-valued Baker-Akhiezer function 
$\Psi(x;\lambda)=\bigl(\Psi^{(ij)}(x;\lambda)\bigr)$ $(i,j=1,2)$ 
solving the following linear problem 
with a spectral parameter $\lambda$ 
which is invariant under the flows of $x^{[\mu]}_n$'s:
\begin{align}
 \bsigma\,\Psi(x;\lambda)=\Psi(x;\lambda)\,\sigma_3,\qquad
  \bL\,\Psi(x;\lambda)=\Psi(x;\lambda)\,\Lambda,
  \qquad
  \Biggl(
  \Lambda\equiv
   \begin{pmatrix}
    \lambda & 0 \cr 0 & \lambda
   \end{pmatrix}
  \Biggr)
 \label{baker2cKP}
\end{align}
and 
\begin{align}
g\,\frac{\del\Psi(x;\lambda)}{\del x_n^{[\mu]}}
=(\bsigma^\mu\bL^n)_+\Psi(x;\lambda). \label{linear-prob}
\end{align}

The linear problem and the Douglas equation 
can be best solved by introducing a $2 \times 2$ matrix-valued 
Sato-Wilson operator $\bW(x;\del)=\bunit_2+\sum_{n\ge 1}w_n(x)\del^{-n}$, 
which satisfies
\begin{align}
 \bsigma=\bW\sigma_3\bW^{-1},\quad \bL=\bW \del \bW^{-1}.
\end{align}
The 2cKP equations \eq{2cKP} are equivalent to 
the so-called Sato equations
\begin{align}
g\,\frac{\del\bW}{\del x_n^{[\mu]}}
  =(\bsigma^\mu\bL^n)_+ \bW-\bW\sigma_3^\mu\del^n
  \,\,\bigl(\,=-(\bsigma^\mu\bL^n)_- \bW \bigr).
 \label{Sato-eq}
\end{align}
Then the Baker-Akhiezer function is solved as 
\begin{align}
 \Psi(x;\lambda)&=\bW(x;\del)
  \cdot \exp\Bigl( g^{-1}\sum_{n\geq1}\sum_{\mu=0,1} 
  \lambda^n x_n^{[\mu]}\sigma_3^\mu \Bigr)\nn\\
 &=\Phi(x;\lambda)\cdot\exp\Bigl( g^{-1}\sum_{n\geq1}\sum_{\mu=0,1} 
  \lambda^n x_n^{[\mu]}\sigma_3^\mu \Bigr) 
 \label{BA}
\end{align}
with $\Phi(x;\lambda)\equiv \bW(x;\del)\bigl|_{\del\to\lambda}$. 
Note that the operator $\bW$ is uniquely determined 
up to the right-multiplication of a diagonal pseudo-differential operator 
with constant coefficients: 
 $\bW\to\bW\cdot e^{\sum_{n< 0}c_n\del^{n}} 
\ (c_n\in \mathbb R\,\bunit_2 +\mathbb R\,\sigma_3)$.

\subsection{Identification with $(p,q)$ minimal superstring theory}
\label{operator_identification}

We now identify the operators of 2cKP hierarchy  
(or the matrix models) with those of super Liouville theory. 
There are two types of minimal superstring theories (see, e.g., \cite{UniCom}):
\begin{itemize}
 \item
 \underline{\bf $(p,q)$ even minimal superstrings}: 
 $(p,q)=(2\hat p,2\hat q)$ 
   with $\hat p+\hat q \in 2\mathbb Z + 1$,
 \item
 \underline{\bf $(p,q)$ odd minimal superstrings}: 
 $(p,q)=(\hat p,\hat q)$ with $\hat p,\,\hat q\in 2\bZ+1$ 
 ($\hat p+\hat q \in 2\mathbb Z $). 
\end{itemize}
Each has a sequence of operators 
$\cO^{[r-s]}_{n} \ (n\equiv \hat qr-\hat ps)$, 
where $r-s\in 2\mathbb Z$ corresponds to NS-NS operators 
and $r-s\in 2\mathbb Z+1$ to R-R operators.%
\footnote{
$[r-s]$ implies $r-s$ modulo $2$.
} 
They are expressed as the gravitational dressing of 
the corresponding operators $\cO^{({\rm M})[r-s]}_n$ 
of $(p,q)$ minimal superconformal matters 
which have the conformal dimensions 
\begin{align}
 \Delta^{({\rm M})[r-s]}_{n=(\hat qr-\hat qs)}
  =\frac{(\hat qr-\hat ps)^2-(\hat q-\hat p)^2}{8\hat q\hat p}
  +\frac{1-(-1)^{r-s}}{32}. 
\label{Kac}
\end{align}
The string susceptibility $\gamma_{\rm str}$ is measured 
with the operator $\cO_{\rm min}$ of minimal scaling dimension 
which is given by either of $\cO^{[0]}_1$ (NS-NS) 
or $\cO^{[1]}_1$ (R-R) for even minimal superstrings 
and by $\cO_{\rm min}=\cO^{[1]}_1$ (R-R) for odd minimal superstrings, 
and is found to be
\begin{align}
 \gamma_{\rm str}=-\frac{1}{\hat p+\hat q-1}
  =\begin{cases}
    -\dfrac{2}{p+q-2} & \mbox{(even)} \\
    -\dfrac{1}{p+q-1} & \mbox{(odd)}
   \end{cases}.
\end{align}
The gravitational scaling dimensions $\Delta^{[\mu]}_n$ of 
the operators $\cO^{[\mu]}_n$ are then defined by 
\begin{align}
 \frac
 {
 \bigl< \cO^{[\mu_1]}_{n_1}\cdots \cO^{[\mu_N]}_{n_N}
  \,e^{-t\,\cO_{\rm min}} \bigr>
  }{
 \bigl< \cO_{\rm min}\cdots \cO_{\rm min}
  \,e^{-t\,\cO_{\rm min}} \bigr>
  }
 \equiv |t|^{\sum_{k=1}^N\Delta^{[\mu_k]}_{n_k}}, 
\end{align}
and given by 
\begin{align}
 \Delta^{[0]}_n=\Delta^{[1]}_n=\frac{\hat p+\hat q-n}{\hat p+\hat q-1}.
\end{align}

To identify the operators, 
we first note that the operators 
$(\bL^n)_+$ and $(\bsigma \bL^n)_+$  
are differential operators of the same order. 
They will give the same gravitational scaling dimensions 
and thus should be considered in pairs. 
Furthermore, according to the analysis of one-matrix model 
with a symmetric double-well potential, 
the excitations symmetric under the reflection belong to the NS-NS sector, 
and those antisymmetric to the R-R sector \cite{tt,newhat,UniCom}. 
Following the derivation given in subsection \ref{2-cut_ansatz},
one can easily see that such reflection is realized in our 2cKP 
by the transformation ${\rm C}:(\bsigma,\bL) \to (-\bsigma,\bL)$.%
\footnote{
This charge conjugation will be studied more in subsection \ref{CCt}.
} 
This implies that in the pair $\bigl((\bL^n)_+,\,(\bsigma \bL^n)_+\bigr)$, 
the former $(\mu=0)$ should belong to the NS-NS sector 
and the latter ($\mu=1$) to the R-R sector. 
This consideration naturally leads us to the following ansatz 
on the operator identification:
\begin{align}
\cO_{n}^{[r-s]} \ \leftrightarrow \ \bigl(\bsigma^{r-s}\bL^n\bigr)_+
 = \sigma_3^{r-s}\,\del^n+\cdots.
\end{align}
See Fig.\ 2 (a) and (b1) for 
the Kac table of the $(p,q)=(2\hat p,2\hat q)=(4,6)$ and 
$(p,q)=(\hat p,\hat q)=(3,5)$ 
minimal superconformal matters.

\begin{figure}[htbp]
\begin{center}
\resizebox{!}{45mm}{
\input{Kac2.pstex_t}}
\end{center}
\caption{\footnotesize{
Kac table for $(p,q)=(4,6)$ and $(p,q)=(3,5)$ minimal conformal matters. 
Letters in boxes denote the corresponding differential operators 
$(\bsigma^\mu\bL^n)_+\leftrightarrow\cO_{\hat qr-\hat ps}^{[r-s]}$. 
The usual conformal block of minimal conformal matters
is shown by a rectangular with bold lines. 
The table (b2) expresses those differential operators of $(3,5)$ model
that do not have their counterparts in Liouville theory. 
They are all odd under the $\bZ_2$ transformation,
${\rm \tilde C}: (\bsigma,\bL)\to(-\bsigma,-\bL)$.
}}
\label{kac-table}
\end{figure}%

We are now in a position to discuss that 
some operators do not appear in the spectrum. 
First, the operators $\bP^n=(\bP^n)_+$ 
do not have their counterparts $\cO^{[n]}_{n\hat p}$ 
in the Kac table of minimal conformal matters. 
However, as can be seen in \eq{deforPQ} or in \eq{2cKP}, 
the evolution along these directions is trivial, 
so that correlation functions including such operators always vanish. 

Second, while the 2cKP hierarchy allows the operators 
$(\bsigma^{n+1}\bL^n)_+$,
they do not have their counterparts in the Kac table 
of $(p,q)$ odd minimal conformal matters (Fig.\ 2 (b2)). 
In this case one can reduce the system 
by orbifolding it with another $\bZ_2$ symmetry, 
$\tilde{\rm C}:(\bsigma,\bL)\to(-\bsigma,-\bL)$, 
after which we have a complete one-to-one correspondence 
between the operators of 2cKP hierarchy 
and those of super Liouville theory. 
We study this orbifolding in detail 
in subsection \ref{CCt}.


We here comment that the spectrum of type 0A $(p,q)$ minimal superstrings 
can be obtained by orbifolding those of type 0B minimal superstrings 
with the $\mathbb Z_2$ transformation ${\rm C}$. 
This is because there are no R-R fields in 0A minimal theories 
other than the degrees of freedom of R-R fluxes \cite{UniCom}. 
We also comment that there exist a series of critical points 
characterized by $\bQ=c_q^{[1]}\del^{\hat q} + \cdots$, 
which can be obtained by perturbing critical systems with R-R operators. 
The $\tilde{\rm C}$ invariance holds  
when $\hat p+\hat q\in 2\bZ+1$ at the new fixed points.

\subsection{Grassmannian of 2cKP hierarchy and free fermions}

In this subsection, we summarize basic properties of 2cKP hierarchy 
and its free-fermion realization.

We introduce another set of evolution parameters 
$x_n^{(i)}$ $(i=1,2)$ 
which are related to $x_n^{[\mu]}$ $(\mu=0,1)$ as 
\begin{align}
 x_n^{(i)}=x_n^{[0]}+(-1)^{i-1} x_n^{[1]}\quad(i=1,2),
\end{align}
or
\begin{align}
 x^{[\mu]}_n=\frac{1}{2}\,\bigl(x^{(1)}_n+(-1)^\mu x^{(2)}_n\bigr)\quad(\mu=0,1).
\end{align}
Recall that $x^{[0]}_n$ (or $x^{[1]}_n$) correspond to the background sources 
for NS-NS (or R-R) operators $\cO^{[0]}_n$ (or $\cO^{[1]}_n$), 
which, in the language of matrix models 
with a symmetric double-well potential, 
describe symmetric (or antisymmetric) excitations of eigenvalues. 
Thus, the operators 
$\cO^{(i)}_n\equiv (1/2)\bigl(\cO_n^{[0]}+(-1)^{i-1}\cO_n^{[1]}\bigr)$ 
with $i=1$ (or $i=2$) describes 
the excitations of the eigenvalues in the left (or right) well.

The Sato equations \eq{Sato-eq} are then expressed as 
\begin{align}
g\,\frac{\del\bW}{\del x_n^{(i)}}
  &=(\be^{(i)}\bL^n)_+\bW-\bW E^{(i)}\del^n\qquad(i=1,2),
\label{Sato-eq2}
\end{align}
where 
\begin{align}
 \be^{(i)}=\bW E^{(i)} \bW^{-1},\qquad
 E^{(1)}=
 \begin{pmatrix}
  1 & 0 \cr 0 & 0
 \end{pmatrix},\quad
 E^{(2)}=
 \begin{pmatrix}
  0 & 0 \cr 0 & 1
 \end{pmatrix}.
\end{align}
For a given solution $\bW(x;\del)$ to the Sato equations, 
we define a series of $2\times 2$ matrix-valued functions of $\lambda$, 
$\Phi_K(x;\lambda)
=\bigl( \Phi_K^{(ij)}(x;\lambda) \bigr)$ $(K=0,1,2,\cdots)$, as
\begin{align}
 \Phi_K(x;\lambda)\equiv e^{-(1/g)x^{[0]}_1 \lambda}\cdot
  \del^K\cdot \bW(x;\del) \cdot e^{(1/g) x^{[0]}_1 \lambda}
  =\bigl[ \del^K\cdot\bW(x;\del)\bigr]\Bigr|_{\del\to\lambda}.
 \label{Phi_K}
\end{align}
Note that $\Phi_{K=0}(x;\lambda)=\Phi(x;\lambda)$ 
(see eq.\ \eq{BA}).

In order to interpret the Sato equations as describing the motion 
in an infinite dimensional Grassmannian $\cM$ \cite{sato-sato}, 
we first introduce an infinite dimensional vector space $\cH$ 
consisting of 2-component row vectors  
$\boldsymbol{v}(\lambda)=(v^{(1)}(\lambda),v^{(2)}(\lambda))$ 
whose components are functions of $\lambda$. 
We also introduce the subspace $\cV_0$ 
which is spanned by the vectors 
$\boldsymbol{v}^{(1)}_K(\lambda)=(\lambda^K,0)$ $(K\geq0)$
and 
$\boldsymbol{v}^{(2)}_K(\lambda)=(0,\lambda^K)$ $(K\geq0)$. 
The infinite dimensional Grassmannian $\cM$ is then defined 
as the set of those subspaces of $\cH$ 
that have the same ``size" with $\cV_0$:%
\footnote{
See, e.g., \cite{sato-sato,SW,djkm2c,KaLe} for a more rigorous definition. 
} 
\begin{align}
 \cM\equiv \bigl\{ \cV\subset \cH \,|\, \cV\sim \cV_0\bigr\}.
\end{align}

For a given Sato-Wilson operator $\bW(x,\del)$ 
and the corresponding sequence of matrices $\Phi_K(x;\lambda)$ 
given in \eq{Phi_K}, 
we introduce a point $\cV(x)\in \cM$ 
as such that is spanned by a series of row vectors 
$\bPhi_K^{(i)}(x;\lambda)\equiv 
\bigl(\Phi_K^{(i1)}(x;\lambda),\,\Phi_K^{(i2)}(x;\lambda)\bigr)$ 
$(K\geq0;\,i=1,2)$.
We denote this by using the $2\times 2$ matrices 
$\Phi_K(x;\lambda)=
\bigl(\bPhi_K^{(1)}(x;\lambda),\bPhi_K^{(2)}(x;\lambda)\bigr)^{\!\rm T}$ 
$(K\geq0)$
as%
\footnote{
$\cV_0$ is then expressed as 
$ \biggl< \biggl(\begin{array}{cc}
  \lambda^K & 0 \cr 0 & \lambda^K \end{array}\biggr)
 \biggr>_{K\geq0}
$
and corresponds to the Sato-Wilson operator $\bW=\bunit_2$.
}
\begin{align}
 \cV(x)\equiv \bigl< \Phi_K(x;\lambda)\bigr>_{K\geq 0}. 
\end{align}
The $x$-evolutions of the linear subspace $\cV(x)$ 
can then be read off by looking at those of the matrices 
$\Phi_K(x;\lambda)$ $(K\geq0)$: 
\begin{align}
 g\,\frac{\del\Phi_K(x;\lambda)}{\del x_n^{(i)}}
  &=-\,\Phi_K(x;\lambda)\,\lambda^n E^{(i)}
  +\Bigl[\del^K\cdot(\be^{(i)}\bL^n){}_+ \cdot\bW\Bigr]\Bigr|_{\del\to\lambda}.
\end{align} 
The second term on the right-hand side 
simply generates a linear transformation among the vectors 
$\bigl\{\bPhi_l^{(i)}(x;\lambda)\bigr\}_{l\leq k,\,i=1,2}$,
and thus can be ignored in considering the motion of $\cV(x)$. 
Integrating this equation, we obtain 
\begin{align}
 \cV(x)=\bigl<\Phi_K(0;\lambda)\,
  e^{-(1/g)\sum_{n,i}x_n^{(i)}\cdot\lambda^n E^{(i)}}
   \bigr>_{K\geq0}
  \equiv \cV(0)\cdot e^{-(1/g)\sum_{n,\mu}x_n^{(i)}\cdot\lambda^n E^{(i)}}.
 \label{motion_in_UGM}
\end{align}
Here $\cV(0)$ is the subspace in $\cH$ 
corresponding to the initial value of $\bW$ at $x=0$.

The free-fermion realization is obtained by making a map from 
this Grassmannian to the Fock space of a free-fermion system. 
Because the linear space $\cV(x)$ is spanned 
by 2-component vector-valued functions $\bPhi_K^{(i)}(x;\lambda)$ 
$(K\geq0;\,i=1,2)$, 
we need two pairs of free chiral 
fermions on the complex $\lambda$ plane, 
$(\psi^{(i)}(\lambda),\bar\psi^{(i)}(\lambda)), \, (i=1,2)$ \cite{djkm2c}.
Then a point $\cV(x)\in\cM$ is assigned with the following state 
$\ket{\Phi(x)}$ in the fermion Fock space:
\begin{align}
 \ket{\Phi(x)}\equiv \prod_{K\geq0}\prod_{i=1,2}
  \Bigl[\sum_{j=1,2}\oint\frac{d\lambda}{2\pi i}\,
  \Phi_K^{(ij)}(x;\lambda)\,
  {\bar\psi}^{(j)}(\lambda)\Bigr]\ket{\Omega}.
 \label{ket1}
\end{align}
The sum over $j\,(=1,2)$ reflects the fact that 
$\bPhi^{(i)}_K=(\Phi_K^{(i1)},\Phi_K^{(i2)})$ is a two-component vector.

We introduce free chiral bosons 
\begin{align}
 \phi^{(i)}(\lambda)
  &=\bar\phi^{(i)}+\alpha^{(i)}_0\ln\lambda
    +\phi^{(i)}_-(\lambda)+\phi^{(i)}_+(\lambda) \nn\\
 &=\bar\phi^{(i)}+\alpha^{(i)}_0\ln\lambda
  -\sum_{n<0} \frac{\alpha^{(i)}_n}{n}\,\lambda^{-n}
  -\sum_{n>0} \frac{\alpha^{(i)}_n}{n}\,\lambda^{-n}
  \quad(i=1,2)
\end{align}
with the OPE 
$\phi^{(i)}(\lambda)\,\phi^{(j)}(\lambda')=\delta^{ij}\,\ln(\lambda-\lambda')$ 
(equivalently, $[\alpha^{(i)}_m,\alpha^{(j)}_n]=m\,\delta^{ij}\,\delta_{m+n,0}$ 
and $[\alpha^{(i)}_0,\bar\phi^{(j)}]=\delta^{ij}$),
to bosonize the free fermions as%
\footnote{
We define 
$\nol e^{\sum_i \beta^{(i)}\phi^{(i)}(\lambda)}\nor
\equiv  e^{\sum_i \beta^{(i)}\phi_-^{(i)}(\lambda)}\,
e^{\sum_i \bar\phi^{(i)}}\,\lambda^{\sum_i \alpha^{(i)}_0}\,
e^{\sum_i\beta^{(i)}\phi_+^{(i)}(\lambda)}.
$
} 
\begin{align}
 \psi^{(i)}(\lambda)
  &=\sum_{r\in\bZ+1/2}\psi^{(i)}_r\lambda^{-r-1/2}
  =\nol e^{\phi^{(i)}(\lambda)}\nor K_i,\\
 \bar\psi^{(i)}(\lambda)
  &=\sum_{r\in\bZ+1/2}\bar\psi^{(i)}_r\lambda^{-r-1/2}
  =\nol e^{-\phi^{(i)}(\lambda)}\nor K_i,
\end{align} 
where $K_i$ is the cocycle, $K_1=1$ and $K_2=(-1)^{\alpha^{(1)}_0}$.
Conversely, the bosons are expressed as 
\begin{align}
 \del\phi^{(i)}(\lambda)\equiv
  \nol\psi^{(i)}(\lambda)\bar\psi^{(i)}(\lambda)\nor 
  =\sum_{n\in\bZ}\alpha^{(i)}_n\lambda^{-n-1}.
\end{align}
Since 
$\bigl[\alpha_n^{(i)},\bar\psi^{(j)}(\lambda)\bigr]=-\delta^{ij}
\lambda^n\,\bar\psi^{(i)}(\lambda)$, 
the motion \eq{motion_in_UGM} of a given point 
in the Grassmannian $\cM$ is expressed as%
\footnote{
The factor $\rho(x)$ reflects the fact 
that the correspondence between a linear space and a fermion state 
is one-to-one only up to a multiplicative factor.
}
\begin{align}
 \ket{\Phi(x)}&= \rho(x)\, 
  e^{+(1/g)\sum_{n\geq1}
  (x_n^{(1)} \alpha_n^{(1)}+ x^{(2)}_n \alpha_n^{(2)} ) }
  \ket{\Phi}.
 \label{ket22}
\end{align}
Here $\ket{\Phi}\equiv\ket{\Phi(0)}$ is the initial state at $x=0$.  
From this state, the Sato-Wilson operator $\bW(x;\del)$ 
can be reconstructed by using the following formula 
for $\Phi(x;\lambda)=\bigl(\Phi^{(ij)}(x;\lambda)\bigr)=\
\bW(x;\del)|_{\del\to\lambda}$: 
\begin{align}
 \Phi^{(ij)}(x;\lambda)
 =\dfrac{
  \bigl<0\bigr|e^{-\bar\phi^{(i)}} \psi^{(j)}(\lambda)\bigl|\Phi(x)\bigr>}
  {\bigl<0\bigl|\Phi(x)\bigr>}
 =\epsilon_{ji}\lambda^{-(1-\delta_{ij})}\,
  \dfrac{
  \bigl<x/g\bigr| U^{i-j} \,
  e^{\phi_+^{(j)}(\lambda)} \bigl|\Phi\bigr>}
  {\bigl<x/g\bigl|\Phi\bigr>},
 \label{tau_to_W}
\end{align}
where $U\equiv e^{\bar\phi^{(1)}-\bar\phi^{(2)}}$, 
and $\epsilon_{ij}=1$ when $i\leq j$ and $\epsilon_{ij}=-1$ when $i>j$. 
A proof of this statement can be found, e.g., in \cite{KaLe}.
Then the Baker-Akhiezer function is obtained as
\begin{align}
 \Psi^{(ij)}(x;\lambda)&=\Phi^{(ij)}(x;\lambda)\cdot
 \exp \bigl(g^{-1}\sum_{n}x^{(j)}_n\lambda^n\bigr) \nn\\
&=\dfrac{
  \bigl<x/g\bigr|e^{-\bar\phi^{(i)}} \,\psi^{(j)}(\lambda)\bigl|\Phi\bigr>}
  {\bigl<x/g\bigl|\Phi\bigr>} . \label{BakerFermion}
\end{align}

We should note that not every state in the fermion Fock space 
can be expressed as a point in the Grassmannian $\cM$. 
This correspondence holds 
if and only if the state $\ket{\Phi}$ is decomposable, 
i.e.\ it can be written 
as $\ket{\Phi}=e^H\ket{0}$ with a fermion bilinear operator $H$. 
Since such $H$ is an element of the Lie algebra $gl(\infty)$ 
realized over the fermion Fock space, 
the Grassmannian is characterized 
as an orbit of the infinite dimensional group $GL(\infty)$ 
\cite{sato-sato}.

\subsection{Background R-R flux}

As is investigated in one-matrix models \cite{BDJT,UniCom,SeSh2}, 
critical points of 2-cut solutions can have a configuration 
where the Fermi levels of two Fermi surfaces 
differ by an integer number $\nu$, 
which is interpreted as background R-R flux \cite{UniCom,SeSh2}. 
In the language of 2cKP hierarchy, 
such configuration is described by a fermion sector 
with $\alpha^{(1)}_0=-\alpha^{(2)}_0=\nu$ 
and can be measured by taking the inner product with the state
\begin{align}
 \bra{x/g;\,\nu}\equiv \bra{x/g} \,U^{-\nu}, 
\end{align}
where $U\equiv e^{\bar\phi^{(1)}-\bar\phi^{(2)}}$. 
Thus, the $\tau$ function 
\begin{align}
 \tau_\nu(x)\equiv \bra{x/g;\nu}\Phi\bigr>
\end{align}
corresponds to the partition function 
to be obtained in a continuum limit 
with fixing the discrepancy of the two Fermi levels 
to be $\nu$. 
The connected correlation functions with background R-R flux are then given 
by setting backgrounds as $x=(b^{[\mu]}_n)$: 
\begin{align}
 \vev{\cO_{\,n_1}^{[\mu_1]}\cdots
  \cO_{\,n_N}^{[\mu_N]}}_{\!\nu,\rm c}
 =\biggl[\frac{\bra{b/g;\nu} \alpha_{\,n_1}^{[\mu_1]}\cdots
  \alpha_{\,n_N}^{[\mu_N]}\ket{\Phi}}
   {\bra{b/g;\nu}\,\Phi\bigr>}\biggr]_{\rm c}
\end{align}
and have an expansion in the string coupling $g$ as
\begin{align}
 \hspace{40mm}=\sum_{h\geq0}\,g^{2h+N-2}\,
  \vev{\cO_{\,n_1}^{[\mu_1]}\cdots
  \cO_{\,n_N}^{[\mu_N]}}^{\!(h)}_{\!\nu,\rm c}.
 \label{g_expansion}
\end{align}
Note that the state $\bra{x/g;\nu}$ is defined   
in the weak coupling region $|\zeta|\to\infty$, 
and thus is a state representing a condensate 
of microscopic-loop operators \cite{Notes_on}. 
\begin{figure}[htbp]
\begin{center}
\resizebox{!}{45mm}{
\input{maya.pstex_t}}
\caption{\footnotesize{
Maya diagram of the state $\ket{\nu}=U^\nu\ket0$.
}}
\label{maya}
\end{center}
\end{figure}%

The Sato-Wilson operator $\bW(x;\del)$ corresponding to 
the $\tau$ function $\tau_\nu(x)$ 
can be reconstructed 
by applying the formula \eq{tau_to_W} to 
$\Phi_\nu(x;\lambda)=\bigl(\Phi^{(ij)}_\nu(x,\nu;\lambda)\bigr)=
\bW(x;\del)|_{\del\to\lambda}$: 
\begin{align}
 \Phi^{(ij)}_\nu(x;\lambda)
 &=\epsilon_{ji}\lambda^{-(1-\delta_{ij})}\,
  \dfrac{
  \bigl<x/g;\nu\bigr| U^{i-j} \,
  e^{\phi_+^{(j)}(\lambda)} \bigl|\Phi\bigr>}
  {\bigl<x/g;\nu\bigl|\Phi\bigr>}\nn\\
 &=\epsilon_{ji}\lambda^{-(1-\delta_{ij})}\,
  \frac{1}{\tau_\nu(x)}\,
  \bigl<x/g;\nu\bigr| U^{i-j} \,
  \exp\Bigl(-\sum_{m\geq1}\,\frac{\alpha^{(j)}_m}{m}
   \lambda^{-m} \Bigr)\bigl|\Phi\bigr>\nn\\
 &=\epsilon_{ji}\lambda^{-(1-\delta_{ij})}\,
  \frac{1}{\tau_\nu(x)}\,
  \exp\Bigl(-g\sum_{m\geq1}\,\frac{\del^{(j)}_m}{m}
   \lambda^{-m} \Bigr)  \bigl<x/g;\nu-i+j\bigr|
   \,\Phi\bigr>\nn\\
 &=\epsilon_{ji}\lambda^{-(1-\delta_{ij})}\,
  \frac{1}{\tau_\nu(x)}\,
  \sum_{n\geq0}\,\lambda^{-n}\,\cS_n\bigl([-g\,\del^{(j)}_m/m]\bigr)\,
  \tau_{\nu-i+j}(x),
 \label{tau_to_W2}
\end{align}
where $\cS_n([y_n])$ are the Schur polynomials defined by 
the generating function, 
$\ds\exp\Bigl[\sum_{m\geq 1}y_m z^m\Bigr]$
$=$ $\sum_{n\in\bZ}\cS_n([y_m])\,z^n$. 
Expanding the left-hand side as 
$\Phi_\nu(x;\lambda)=1+\sum_{n\geq1}w_n(x)\,\lambda^{-n}$ 
and comparing the coefficients of $\lambda^{-n}$ $(n=1,2,\cdots)$, 
we obtain the formula
\begin{align}
 w_n(x)=\begin{pmatrix}
  \dfrac{
  \cS_n\bigl([-g\,\del^{(1)}_m/m]\bigr)\,\tau_\nu(x)}{\tau_\nu(x)}
  &
  -\,\dfrac{
  \cS_{n-1}\bigl([-g\,\del^{(2)}_m/m]\bigr)\,\tau_{\nu+1}(x)}{\tau_\nu(x)}
  \\
  \dfrac{
  \cS_{n-1}\bigl([-g\,\del^{(1)}_m/m]\bigr)\,\tau_{\nu-1}(x)}{\tau_\nu(x)}
  &
  \dfrac{
  \cS_n\bigl([-g\,\del^{(2)}_m/m]\bigr)\,\tau_\nu(x)}{\tau_\nu(x)}
 \end{pmatrix}.
 \label{rel1}
\end{align}
The first two are then given by
\begin{align}
 w_1(x)=&
  \begin{pmatrix}
  -g\,\del_1^{(1)}\ln \tau_\nu & -\dfrac{\tau_{\nu+1}}{\tau_\nu} \cr 
  \dfrac{\tau_{\nu-1}}{\tau_\nu} & -g\,\del_1^{(2)}\ln \tau_\nu
  \end{pmatrix}, 
 \label{rel1a}\\
 w_2(x)=&
  \begin{pmatrix}
  \dfrac{\bigl(-g\,\del_2^{(1)}+(g\,\del_1^{(1)})^2\bigr)\tau_\nu}{2\tau_\nu} 
  & \dfrac{g\,\del_1^{(2)}\tau_{\nu+1}}{\tau_\nu} \cr 
  -\,\dfrac{g\,\del_1^{(1)}\tau_{\nu-1}}{\tau_\nu} 
  & \dfrac{\bigl(-g\,\del_2^{(2)}+(g\,\del_1^{(2)})^2\bigr)\tau_\nu}{2\tau_\nu} 
  \end{pmatrix}.
 \label{rel1b}
\end{align}
On the other hand, expanding in $\del$ the Sato equations 
\begin{align}
 g\,\del_n^{(i)} \bW (x;\del)=-(\be^{(i)}\bL^n)_-\bW(x;\del),
\end{align}
we obtain the relation between $(\be^{(i)}\bL^n)_{-k}(x)$ and $w_n(x)$ 
$\bigl($defined by 
$ \bigl(\be^{(i)}\bL^n\bigr)(x;\del)$
$=$ $\sum_{k\in \mathbb Z}$ $\bigl(\be^{(i)}\bL^n\bigr)_{k}(x)$ $\del^{k}$ 
and $\bW(x;\del) =\sum_{n\ge 0}w_n(x)\,\del^{-n}\bigr)$: 
\begin{align}
 (\be^{(i)}\bL^n)_{-1}=-g\,\del_n^{(i)}w_1(x), \qquad
  (\be^{(i)}\bL^n)_{-2}=-g\,\del_n^{(i)}w_2(x)
   +g\,\del_n^{(i)}w_1(x)\cdot w_1(x),
  \quad\cdots
 \label{rel2}
\end{align}
Equations \eq{rel1} and \eq{rel2} give the relations 
which express the coefficients of the pseudo-differential operator 
$(\be^{(i)}\bL^n)_-$ in terms of $\tau$ functions.
In particular, looking at the $(1,1)$ component of the relation 
$(\be^{(2)}\bL)_{-1}=-g\,\del^{(2)}_1 w_1(x)$, 
we obtain the formula
\begin{align}
 g^2\,\del^{(1)}_1\del^{(2)}_1\ln\tau_\nu
  \,\,\Bigl(=\frac{g^2}{4}\,\bigl[\bigl(\del^{[0]}_1\bigr)^2
    -\bigl(\del^{[1]}_1\bigr)^2\bigr]\ln\tau_\nu(x)\Bigr)
  =\frac{\tau_{\nu+1}(x)\,\tau_{\nu-1}(x)}{\tau_\nu^2(x)}.
 \label{toda}
\end{align}

\section{$W_{1+\infty}$ constraints and string equations}

In this section, we solve the Douglas equation \eq{douglas-eq}, 
and show that the generating functions (or $\tau$ functions) 
obey the $W_{1+\infty}$ constraints. 
Using this, we  derive string equations in a generic form. 

\subsection{Solutions to the Douglas equation \eq{douglas-eq}} 
\label{SolnDouglas}

To solve the Douglas equation \eq{douglas-eq}  
we first note that $\bP=\bsigma\bL^{\hat p}$ is given by
\begin{align}
\bP=\bW\, \sigma_3 \,\del^{\hat p}\,\bW^{-1}.
\end{align} 
Then the general solution to the Douglas equation \eq{douglas-eq} 
is given by%
\footnote{
For a generic value of $(\theta_p,\theta_q)$ and $\epsilon$, 
we have
\begin{align}
 \bP=\bW\, \sigma_3^{\epsilon}e^{\theta_p\sigma_3}\del^{\hat p}\,\bW^{-1},
 \qquad
 \bQ=\bW\, \frac{1}{\hat p}\,\sigma_3^{\epsilon}e^{-\theta_p\sigma_3}
  \biggl(\,
  \sum_{n\ge 1}\sum_{\mu=0,1}
  n\,x_n^{[\mu]}\, \sigma_3^{\mu}\del^{n-\hat p}+g\gamma \,\del^{-\hat p}
 \biggr)\, \bW^{-1}. \nn
\end{align}
} 
\begin{align}
\bQ=\bW\, \frac{1}{\hat p}\,\sigma_3\,
 \biggl(\,
  \sum_{n\ge 1}\sum_{\mu=0,1}
   n\,x_n^{[\mu]}\sigma_3^{\mu}\del^{n-\hat p}+g\gamma \,\del^{-\hat p}
 \biggr)\, \bW^{-1}
\end{align}
with $\gamma$ being an arbitrary constant to be fixed later.
This can be derived in the way essentially the same with 
that of the bosonic case (\cite{Krichever:1992sw,fkn3}, 
see also Theorem 1 in \cite{fim}).
Then by requiring 
that $\bP$ and $\bQ$ at the initial time $x=(b_n^{[\mu]})$ 
be differential operators of order $\hat p$ and $\hat q$, respectively,  
we set the background as 
$b_n^{[\mu]}=0$ $(n>\hat p+\hat q)$ 
to obtain:
\begin{align}
 \bP&=\bW\, \sigma_3\,\del^{\hat p}\,\bW^{-1}
  = \bigl[\bW\, \sigma_3\,\del^{\hat p}\,\bW^{-1} \bigr]_+, \\
 \bQ&=\bW\, \frac{1}{\hat p}\,\sigma_3\,
  \biggl(\,
  \sum_{n=1}^{\hat p+\hat q}\sum_{\mu=0,1}
  n\,b_n^{[\mu]}\, \sigma_3^{\mu}\del^{n-\hat p}+g\gamma \,\del^{-p}
  \biggr)\, \bW^{-1}\nn\\
 &=\biggl[\bW\, \frac{1}{\hat p}\,
  \biggl(\,
  \sum_{n=\hat p}^{\hat p+\hat q}\sum_{\mu=0,1}
  n\,b_n^{[\mu]}\, \sigma_3^{\mu+1}\del^{n-\hat p}
  \biggr)\, \bW^{-1} \biggr]_+.
\end{align} 
Here we have set $\bW=\bW(b;\del)$. 
Recall that for odd minimal superstrings, 
we need to set the backgrounds as
\begin{align}
 b^{[n+1]}_n=0\quad 
 \mbox{for odd minimal superstrings: 
      $(p,q)=(\hat p,\hat q)$ ($\hat p,\,\hat q\in 2\bZ+1$)}
\end{align}
in order to make $\bQ$ invariant under the $\bZ_2$ transformation 
$\tilde{\rm C}:(\bsigma,\bL)\to(-\bsigma,-\bL)$. 

The action of $\bQ$ on the Baker-Akhiezer function 
in a general background is now calculated as follows:
\begin{align}
 &\bQ\,\Psi(x;\lambda)= \nn\\
  &=\bW\frac{1}{\hat p}\,
  \biggl(\,
  x^{[0]}_1\,\del^{1-\hat p} + x^{[1]}_1\sigma_3\,\del^{1-\hat p}
  + \sum_{n\geq2}\sum_\mu
   n\,x_n^{[\mu]} \sigma_3^{\mu}\,\del^{n-\hat p} + g\gamma \,\del^{-\hat p}
  \biggr)\,\sigma_3\,
  \bW^{-1}\Psi(x;\lambda) \nn\\
 &=\frac{1}{\hat p}\,\bigl[\bW,x^{[0]}_1\bigr]
  \del^{1-\hat p}\sigma_3\bW^{-1}\Psi(x;\lambda)
  +\frac{1}{\hat p}\,\biggl(\,
   \sum_{n\geq1}\sum_\mu
   n\,x_n^{[\mu]} \bsigma^{\mu}\,\bL^{n-\hat p} + g\gamma \,\bL^{-\hat p}
  \biggr)\bsigma\,\Psi(x;\lambda) \nn\\
 &=\bigg\{\,\bigl[\bW(x;\del),x_1^{[0]}\bigr]\cdot\bW^{-1}\Psi(x;\lambda)
  +\Psi(x;\lambda)
 \biggl(\,\sum_{n\geq1}\sum_\mu n\,x_n^\mu\sigma_3^\mu\lambda^{n-1}
  +g\gamma\lambda^{-1}\Bigr)
\biggr\}\frac{\sigma_3}{\hat p}\lambda^{1-\hat p}.
\end{align}
Since $\bigl[\bW,x^{[0]}_1\bigr]
=\bigl[\sum_{n\geq0}w_n(x)\del^{-n},x^{[0]}_1\bigr]
=g\sum_{n\geq1}(-n)\,w_n(x)\,\del^{-n-1}
=g\,\bW(x;\lambda)\,\overleftarrow{\del_\lambda}\Bigr|_{\lambda\to\del}$
$=g\,\Phi(x;\lambda)\,\overleftarrow{\del_\lambda}$, 
the above equation can be further rewritten into the following form:
\begin{align}
 \bQ\,&\Psi(x;\lambda)\nn\\
 &=g\,\bigg\{ 
  \Phi(x;\lambda)\,\overleftarrow{\del_\lambda}\Bigr|_{\lambda\to\del}
  \cdot\bW^{-1}\Psi(x;\lambda)
  +\Phi(x;\lambda)\cdot
  \biggl[\bigl(\bW^{-1}\Psi(x;\lambda)\bigr)\,
  \frac{\overleftarrow{\del}\,}{\del\lambda}\biggr]
  +\gamma \lambda^{-1}
  \biggr\}\frac{\sigma_3}{\hat p}\,\lambda^{1-\hat p}\qquad\qquad\nn\\
 &=g\,\Psi(x;\lambda)\biggl[\,\frac{\overleftarrow{\del}\,}{\del \lambda}
  + \gamma \lambda^{-1}\biggr]
    \begin{pmatrix}
     d\lambda/d\zeta & 0 \cr 0 & -d\lambda/d\zeta
    \end{pmatrix}
  \qquad\bigl(\zeta=\lambda^{\hat p}\bigr).
\end{align}
Setting $\gamma\equiv -(\hat p -1)\nu$, we thus obtain%
\footnote{Note that this is a 
representation of the Douglas equation 
$[\bP,\bQ]=g\bunit_2$ given 
by a right multiplication $[\zeta,\overleftarrow{\del_\zeta}]=1$.} 
\begin{align}
\bP\,\tilde\Psi(x;\zeta)=\tilde\Psi(x;\zeta)\, 
 \begin{pmatrix}
     \zeta & 0 \cr 0 & -\zeta
    \end{pmatrix},\quad 
\bQ\,\tilde\Psi(x;\zeta)=g\,\tilde\Psi(x;\zeta)
\,\begin{pmatrix}
     \overleftarrow{\frac{\del}{\del\zeta}} & 0 
     \cr 0 & -\,\overleftarrow{\frac{\del}{\del\zeta}}
    \end{pmatrix}
\label{wave-baker}
\end{align}
for
\begin{align}
 \tilde\Psi(x;\zeta)\equiv\Psi(x;\lambda)
 \biggl(\frac{\del\lambda}{\del \zeta}\biggr)^{\nu}.
\end{align}
It turns out to be convenient 
to introduce another set of chiral fermions 
$(c_0^{(i)}(\zeta),\bar c_0^{(i)}(\zeta))$ $(i=1,2)$: 
\begin{align}
 c_0^{(i)}(\zeta)\equiv 
  \Bigl(\frac{d\lambda}{d\zeta}\Bigr)^{\!\nu}\,\psi^{(i)}(\lambda),
  \qquad
  \bar c_0^{(i)}(\zeta)\equiv
  \Bigl(\frac{d\lambda}{d\zeta}\Bigr)^{\!\nu}\,\bar\psi^{(i)}(\lambda).
\end{align} 
By using the representation \eq{BakerFermion}, 
$\tilde\Psi(x;\zeta)$ can be written as
\begin{align}
\tilde\Psi^{(ij)}(x;\zeta)=\dfrac{
  \bigl<x/g\bigr|e^{-\bar\phi^{(i)}}c^{(j)}_0(\zeta)\bigl|\Phi\bigr>}
  {\bigl<x/g\bigl|\Phi\bigr>}.
\end{align}
Note that these new fermions have a $\bZ_{\hat p}$ monodromy, 
and thus we introduce:
\begin{align}
 c^{(i)}_a(\zeta)\equiv c^{(i)}_0(e^{2\pi ia}\zeta),\quad
  \bar c^{(i)}_a(\zeta)\equiv \bar c^{(i)}_0(e^{2\pi ia}\zeta)
  \quad(a=0,1,\cdots,\hat p-1).
 \label{monodromy}
\end{align}

\subsection{$W_{1+\infty}$ constraints}\label{Winf_constr}

The corresponding $W_{1+\infty}$ constraints can be derived also in 
the way same with that of bosonic case 
(\cite{Krichever:1992sw,fkn3}, see also 
Lemma 1 in \cite{fim}), 
starting from the following expression of the differential operators 
$\bP$ and $\bQ$:
\begin{align}
\bP\,\bW=\bW\, \sigma_3\,\del^{\hat p},\qquad
\bQ\,\bW=\bW\, \sigma_3
\Bigl[\sum_{n,\mu}n\,x_n^{[\mu]}\sigma_3^\mu\,\del^{n-\hat p}
+g\gamma\del^{-\hat p}\Bigr].
\end{align}
Denoting the initial subspace by 
$\cV\equiv\cV(0)=\cV(x)\cdot 
e^{(1/g)\sum_{n,\mu}x_n^{[\mu]}\sigma_3^\mu\lambda^n}\subset \cH$, 
one can easily show that
\begin{align}
 \cP\equiv \zeta \sigma_3,\qquad
 \cQ\equiv \Bigl[\frac{\overleftarrow{\del}}{\del \lambda}\lambda^{1-\hat p}
 +\gamma\lambda^{-\hat p}\Bigr] \sigma_3
 \quad(\gamma=-(\hat p-1)\nu)
\end{align}
do not leave the vector space $\cV$:
\begin{align}
 \cV\cdot \cP \subset \cV,\qquad
  \cV\cdot \cQ \subset \cV.
 \label{WinfGrassmann0}
\end{align}
Recursively using this invariance, 
we find that 
$\cV$ is invariant under the right action 
of $\cQ^l\,\cP^m$ for arbitrary nonnegative integers $l$ and $m$,

Since $\cQ$ can be written as 
$\ds \Bigl[\Bigl(\frac{d\lambda}{d\zeta}\Bigr)^{\nu}
  \frac{\overleftarrow{\del}}{\del \zeta}
  \Bigl(\frac{d\lambda}{d\zeta}\Bigr)^{-\nu}\Bigr] 
  \sigma_3$,
the above can be restated for the space of functions 
of $\zeta=\lambda^{\hat p}$ with an extra factor 
$(d\lambda/d\zeta)^\nu$:
\begin{align}
 \tilde\cV\equiv \cV\cdot \Bigl(\frac{d\lambda}{d\zeta}\Bigr)^{\nu}
\end{align}
as
\begin{align}
 \tilde\cV\cdot \tilde\cQ^l\,\tilde\cP^m 
  \subset \tilde\cV
  \quad(l,m\geq0)
 \label{WinfGrassmann}
\end{align}
with 
\begin{align}
 \tilde\cP\equiv \zeta \sigma_3,\qquad
 \tilde\cQ\equiv \frac{\overleftarrow{\del}}{\del \zeta}
  \sigma_3.
\end{align}

In term of free fermions, the invariance of the space of functions 
$\cV$ (or $\tilde\cV$) implies the invariance 
of the fermion state $\ket{\Phi}$ 
under the action of the corresponding bilinear operators
\begin{align}
 O\bigl[\cQ^l\,\cP^m\bigr]
  &=\oint\frac{d\lambda}{2\pi i}\,
  \nol\psi(\lambda)\cdot\cQ^l\,\cP^m\cdot \bar\psi(\lambda)\nor \nn\\
  &=\gint\frac{d\zeta}{2\pi i}\,
  :\! c_0(\zeta)\cdot\tilde\cQ^l\,\tilde\cP^m\cdot 
  \bar c_0(\zeta)\!: \nn\\
 & =\gint\frac{d\zeta}{2\pi i}\,
 :\!\Bigl(c_0(\zeta)\frac{\overleftarrow{\del^l}}{\del \zeta^l}\Bigl)
 \zeta^m 
 (\sigma_3)^{m+l}
 \bar c_0(\zeta)\!: \nn\\
 &= \oint\!\frac{d\zeta}{2\pi i}\,\sum_{a=0}^{\hat p-1}
 :\!\Bigl(c_a(\zeta)\frac{\overleftarrow{\del^l}}{\del\zeta^l}\Bigl)
 \zeta^m
 (\sigma_3)^{m+l}
 \bar c_a(\zeta)\!:.
 \label{Winf_lm}
\end{align}
Here the contour of the integral $\ds\gint$ surrounds 
$\zeta=\infty$ $\hat p$ times.  
We further bosonize the fields 
\begin{align}
 c_a(\zeta)=\bigl(c_a^{(1)}(\zeta),c_a^{(2)}(\zeta)\bigr),\qquad
 \bar c_a(\zeta)=\binom{\bar c_a^{(1)}(\zeta)}{\bar c_a^{(2)}(\zeta)},
 \qquad(a=0,1,\cdots,\hat p-1)
\end{align}
as%
\footnote{
$K^{(i)}_a$ are proper cocycles which ensure the anticommutation 
relations among fermion fields with different indices,
They can be taken, for example, as 
$\ds K^{(i)}_a=\prod_{j=1}^{i-1}\prod_{b=0}^{a-1}(-1)^{\alpha^{(j)}_{b,0}}$
with $\alpha^{(i)}_{a,0}$ are the momentum of $\varphi^{(i)}_a(\zeta)$. 
} 
\begin{align}
 c_a^{(i)}(\zeta)
  =\,:e^{\varphi^{(i)}_a(\zeta)}:K^{(i)}_a,\qquad
 \bar c_a^{(i)}(\zeta)
  =\,:e^{-\varphi^{(i)}_a(\zeta)}:K^{(i)}_a
\end{align}
with two sets of free chiral bosons
\begin{align}
 \varphi^{(i)}_a(\zeta)\,\varphi^{(j)}_b(\zeta')
  = +\,\delta^{ij}\,\delta_{ab}\,\ln(\zeta-\zeta')\quad
  \bigl(i,j=1,2;~a,b=0,1,\cdots,\hat p-1\bigr). 
\end{align}
The normal ordering $:\,\,:$ is now the one 
taken with respect to the $SL(2,\bC)$ 
invariant vacuum for $\zeta$ plane, $\ket{0}_{\!\zeta}$, 
so that the original vacuum $\ket{0}$ 
(respecting the $SL(2,\bC)$ invariance for $\lambda$) 
is interpreted as the twisted vacuum for the chiral bosons 
living on the $\zeta$ plane.
The monodromy \eq{monodromy} now should be understood 
as relations to hold in correlation functions 
with this twisted vacuum (not as operator identities).

By writing $l=s-1$ and $m=s+n-1$ $(s\geq1;\,n\geq -s+1)$ 
in \eq{Winf_lm}, 
the operator $O[\cQ^l\,\cP^m]=O[\cQ^{s-1}\,\cP^{s+n-1}]$ is then written as%
\footnote{
One can show that the generic case of \eq{generic} 
can be recovered by properly changing the integration variables 
$\zeta$ separately for $\varphi^{(1)}(\zeta)$ and $\varphi^{(2)}(\zeta)$. 
\label{generic2}} 
\begin{align}
 &\frac{1}{s}\,\gint\frac{d\zeta}{2\pi i} \,\zeta^{s+n-1}\,\sum_a\,\Bigl[
  :\!e^{-\varphi^{(1)}(\zeta)}\del_\zeta^s e^{\varphi^{(1)}(\zeta)}\!:
  +(-1)^n :\!e^{-\varphi^{(2)}(\zeta)}\del_\zeta^s e^{\varphi^{(2)}(\zeta)}
  \!:\Bigr]\nn\\
 &~=\frac{1}{s}\,\gint\frac{d\zeta}{2\pi i} \,\zeta^{s+n-1}\,\sum_a\,\Bigl[
  :\!e^{-\varphi^{(1)}(\zeta)}\del_\zeta^s e^{\varphi^{(1)}(\zeta)}\!:
  +:\!e^{-\varphi^{(2)}(-\zeta)}\del_\zeta^s e^{\varphi^{(2)}(-\zeta)}
  \!:\Bigr]. 
\end{align}
Thus, if we define the $W_{1+\infty}$ currents 
$W^s(\zeta)$ $(s=1,2,\cdots)$ as
\begin{align}
 W^s(\zeta)
 &\equiv \sum_{a=0}^{\hat p-1}\,\cW^s_a(\zeta)\nn\\
 &=\sum_{a=0}^{\hat p-1}\,
  \bigl( 
   :e^{-\varphi^{(1)}_a(\zeta)}\del^s\,e^{\varphi^{(1)}_a(\zeta)}:
  +:e^{-\varphi^{(2)}_a(-\zeta)}\del^s\,e^{\varphi^{(s)}_a(-\zeta)}:
  \bigr)\nn\\
 &\equiv\sum_{n\in\bZ}\,W^s_n\,\zeta^{-n-s}
 \label{Winf_def}
\end{align}
with the coefficient modes $W^s_n$ 
that naturally constitute the $W_{1+\infty}$ algebra \cite{Winf}, 
then eq.\ \eq{WinfGrassmann} implies the following 
$W_{1+\infty}$ constraints 
on the state $\ket{\Phi}$
\begin{align}
 W^s_n\ket{\Phi}={\rm const.}\ket{\Phi} \qquad (s\geq1;\,n\geq-s+1).
\end{align}
One can show that any element in the Borel subalgebra 
spanned by $W^s_{n}$ with $s\geq1$ and $n\geq -s+1$ 
can always be written as a commutator 
of other two elements belonging to the same subalgebra 
(see, e.g., Lemma 4.2 of \cite{fkn2}), 
and thus the constants should vanish. 
We thus obtain the  $W_{1+\infty}$ constraints on the state $\ket{\Phi}$:
\begin{align}
 W^s_n\ket{\Phi}=0\qquad (s\geq1;\,n\geq-s+1).
 \label{Winf}
\end{align}
This is equivalent to the statement 
that the action of the $W_{1+\infty}$ currents on the state $\ket{\Phi}$ 
only yields regular dependence on $\zeta$:
\begin{align}
 W^s(\zeta)\,\ket{\Phi}:\,\mbox{regular around $\zeta=0$}.
 \label{Winf_zeta}
\end{align}
This is the key relation to determine loop amplitudes.

The first two of the $W_{1+\infty}$ currents are given by
\begin{align}
 W^1(\zeta)&=\sum_a\,\bigl( 
  \del\varphi_a^{(1)}(\zeta)-\del\varphi_a^{(2)}(-\zeta)
  \bigr),\\
 W^2(\zeta)&=
  \sum_{a}\Bigl(
   \,:\!\bigl(\del\varphi_a^{(1)}(\zeta)\bigr)^2\!:
    +\del^2\varphi_a^{(1)}(\zeta)
   +:\!\bigl(-\del\varphi_a^{(2)}(-\zeta)\bigr)^2\!:
    +\del^2\varphi_a^{(2)}(-\zeta)
  \Bigr)\nn\\
 &=\sum_{a}\Bigl(
   \nol\bigl(\del\varphi_a^{(1)}(\zeta)\bigr)^2\nor
    +\del^2\varphi_a^{(1)}(\zeta)
   +\nol\bigl(-\del\varphi_a^{(2)}(-\zeta)\bigr)^2\nor
    +\del^2\varphi_a^{(2)}(-\zeta)
  \Bigr)
  +\frac{\hat p^2-1}{6\hat p}\frac{1}{\zeta^2}\nn\\
 &\equiv 2 \sum_{n\in\bZ}\,L_n\,\zeta^{-n-2}+\del W^1(\zeta).
\end{align}
Since the normal ordering with respect to the original vacuum $\ket{0}$ 
can be represented in terms of oscillators, 
we have
\begin{align}
 W^1_n&=\alpha^{(1)}_{n\hat p}+(-1)^n\alpha^{(2)}_{n\hat p},\\
 L_n&=\frac{1}{2\hat p}\sum_{n\in\bZ}\Bigl(
  \nol\alpha^{(1)}_{n\hat p-m}\alpha^{(1)}_m\nor
  +(-1)^n\nol\alpha^{(2)}_{n\hat p-m}\alpha^{(2)}_m\nor\Bigr)
  +\frac{\hat p^2-1}{12\hat p}\,\delta_{n,0}.
\end{align}
These imply 
the $p^{\rm th}$ reduction conditions and the Virasoro constraints 
on the state (as in the bosonic case \cite{fkn1,dvv,gn,g}): 
\begin{align}
 \bigl(\alpha^{(1)}_{n\hat p}+(-1)^n\alpha^{(2)}_{n\hat p}\bigr)
  \ket{\Phi}&=0\quad(n\geq0),
 \label{p-th0}\\
 L_n\,\ket{\Phi}&=0\quad(n\geq -1),
 \label{Virasoro0}
\end{align}
which are rewritten for the $\tau$ functions with background R-R flux $\nu$, 
$\tau_{\nu}(x)=\bra{x/g}U^{-\nu}\ket{\Phi}$, as
\begin{align}
 \bigl(\del^{(1)}_{n\hat p}+(-1)^n\del^{(2)}_{n\hat p}\bigr)\,\tau_\nu(x)
  =0,\qquad
 \nu^{(1)}=-\nu^{(2)}=\nu,
 \label{p-th}
\end{align}
and
\begin{align}
 \cL_n\,\tau_\nu(x)=0\quad(n\geq -1) 
 \label{Virasoro}
\end{align}
with
\begin{align}
 \hat p \cL_{+n}&=\frac{g^2}{2}\,\sum_{m=1}^{n\hat p-1}\bigl(
  \del^{(1)}_{n\hat p-m}\,\del^{(1)}_m 
  +(-1)^n \del^{(2)}_{n\hat p-m}\,\del^{(2)}_m \bigr)\nn\\
 &~~~+\sum_{m\geq1}\,m\,
   \bigl(x^{(1)}_{m}\,\del^{(1)}_{m+n\hat p}
   +(-1)^n\,x^{(2)}_{m}\,\del^{(2)}_{m+n\hat p}\bigr)
  +g\nu\bigl(\del^{(1)}_{n\hat p}-(-1)^n\,\del^{(2)}_{n\hat p}\bigr),\\
 \hat p \cL_0&=\sum_{m\geq1}m\,\bigl( x^{(1)}_m\,\del^{(1)}_m
  + x^{(2)}_m\,\del^{(2)}_m \bigr) 
  +\frac{\hat p^2-1}{12} + \nu^2,\\
 \hat p\cL_{-n}&=
  \frac{1}{2g^2}\,\sum_{m=1}^{n\hat p-1}\,m(n\hat p-m)\,
   \bigl(x^{(1)}_{n\hat p-m}\,x^{(1)}_m
   +(-1)^n\,x^{(2)}_{n\hat p-m}\,x^{(2)}_m\bigr)\nn\\
 &~~~+\sum_{m\geq1}\,(n\hat p+m)\,
   \bigl(x^{(1)}_{n\hat p+m}\,\del^{(1)}_m
   +(-1)^n\,x^{(2)}_{n\hat p+m}\,\del^{(2)}_m\bigr)
  +\frac{n\hat p}{g}\,\nu\,\bigl(x^{(1)}_{n\hat p}-(-1)^n\,x^{(2)}_{n\hat p}\bigr).
\end{align}
The differential operators satisfy the Virasoro algebra 
of central charge $2\hat p$, 
$\bigl[\cL_n,\cL_m\bigr]
=(n-m)\,\cL_{n+m}+(2\hat p\,(n^3-n)/12)\,\delta_{n+m,0}$.

Note that the whole $W_{1+\infty}$ constraints are derived 
from two of the constraints $W^1_1\tau_\nu=0$ and 
$W^2_{-1}\tau_\nu=0$ (or equivalently, $\cL_{-1}\tau_\nu=0$) 
if we require that the function $\tau_\nu$ be a $\tau$ function of 2cKP. 
Thus, for a $\tau$ function $\tau_\nu(x)$, 
the equation
\begin{align}
 \cL_{-1}\,\tau_\nu(x)=0
 \label{string_eq0}
\end{align}
together with the $p^{\rm th}$ reduction conditions \eq{p-th} 
has the maximal information to determine the partition function 
with the R-R flux $\nu$.

In terms of pseudo-differential operators, 
eq.\ \eq{string_eq0} is expressed as follows:

\noindent
\underline{$\hat p=1$}:
\begin{align}
 0&=\sum_{n\geq 0}\,(n+1)\,\Bigl[
  x^{(1)}_{n+1}\,\bigl(\be^{(1)}\bL^n\bigr)_{-1}
   - x^{(2)}_{n+1}\,\bigl(\be^{(2)}\bL^n\bigr)_{-1} \Bigr]
   +g\,\nu\,\bunit_2\nn\\
 &=\sum_{n\geq 0}\sum_{\mu=0,1}\,(n+1)\,
  x^{[\mu]}_{n+1}\,\bigl(\bsigma^{\mu+1}\bL^n\bigr)_{-1}
   +g\,\nu\,\bunit_2,
 \label{se_p1}
\end{align}

\noindent
\underline{$\hat p=2$}:
\begin{align}
 0&=\sum_{n\geq -1}\,(n+2)\,\Bigl[
  x^{(1)}_{n+1}\,\bigl(\be^{(1)}\bL^n\bigr)_{-1}
   - x^{(2)}_{n+1}\,\bigl(\be^{(2)}\bL^n\bigr)_{-1} \Bigr]\nn\\
 &=\sum_{n\geq -1}\sum_{\mu=0,1}\,(n+2)\,
  x^{[\mu]}_{n+2}\,\bigl(\bsigma^{\mu+1}\bL^n\bigr)_{-1},
 \label{se_p2a}
 \\
 0&=\sum_{n\geq -1}\,(n+2)\,\Bigl[
  x^{(1)}_{n+2}\,\bigl(\be^{(1)}\bL^n\bigr)_{-2}
   - x^{(2)}_{n+2}\,\bigl(\be^{(2)}\bL^n\bigr)_{-2} \Bigr]
   +g\begin{pmatrix} \nu-1/2 & 0 \cr 0 & \nu+1/2 \end{pmatrix}\nn\\
 &=\sum_{n\geq -1}\sum_{\mu=0,1}\,(n+2)\,
  x^{[\mu]}_{n+2}\,\bigl(\bsigma^{\mu+1}\bL^n\bigr)_{-2}
   +g\begin{pmatrix} \nu-1/2 & 0 \cr 0 & \nu+1/2 \end{pmatrix},
 \label{se_p2b}
\end{align}
where
\begin{align}
 &\bigl(\be^{(1)}\bigr)_{-1}=-\bigl(\be^{(2)}\bigr)_{-1}
  =\begin{pmatrix}0 & H_+ \cr H_- & 0\end{pmatrix},
 \label{form_e1}\\
 &\bigl(\be^{(1)}\bigr)_{-2}=-\bigl(\be^{(2)}\bigr)_{-2}
  =-\begin{pmatrix}H_+ H_- & g\,\del^{(2)}_1H_+ \cr 
                   g\,\del^{(1)}_1H_- & -H_+ H_-\end{pmatrix},
 \label{form_e2}\\
 &\bigl(\be^{(1)}\bL^{-1}\bigr)_{-1}=E^{(1)}
  =\begin{pmatrix}1 & 0 \cr 0 & 0\end{pmatrix},\quad
 \bigl(\be^{(2)}\bL^{-1}\bigr)_{-1}=E^{(2)}
  =\begin{pmatrix}0 & 0 \cr 0 & 1\end{pmatrix},
 \label{form_eL1}\\
 &\bigl(\be^{(1)}\bL^{-1}\bigr)_{-2}=-\bigl(\be^{(2)}\bL^{-1}\bigr)_{-2}
  =\begin{pmatrix}0 & H_+ \cr H_- & 0 \end{pmatrix}
 \label{form_eL2}
\end{align}
with 
\begin{align}
 H_{\pm}(x)\equiv \frac{\tau_{\nu\pm1}(x)}{\tau_\nu(x)}.
\end{align}
A proof of the above formulas is given in Appendix \ref{string_eqn}.

As an example, we consider the $\hat p=1$ case  
with the differential operator 
\begin{align}
 \bP&=\bsigma\bL=\sigma_3\,\del+2H,
 \quad
 H\equiv
  \begin{pmatrix}
  0 & H_+ \cr H_- & 0
  \end{pmatrix}.
\end{align}
The variables $x^{[n-1]}_n$ are only physically important 
due to the $\hat p=1^{\rm st}$ conditions, 
$\del^{[n]}_n\tau_\nu(x)=0$. 
By denoting them by $t_n\equiv x^{[n-1]}_n$,  
the string equations become 
\begin{align}
 \sum_{n=1}^{\hat q+1}\,n\,t_n\,(\bsigma\bP^{n-1})_{-1} 
  + g\,\nu\,\bunit_2=0.
  \label{string_eq_p1}
\end{align}
The pseudo-differential operator $\bsigma$ is easily found 
by requiring $[\bsigma,\bP]=0$ and $\bsigma^2=\bunit_2$, 
and we obtain
\begin{align}
 \bsigma=\sigma_3+2H_1\del^{-1}+2H_2\del^{-2}+2H_3\del^{-3}+\cdots
\end{align}
with the coefficient functions $H_n$ given in \eq{H_n} of Appendix \ref{ZS}.

The string equations for the backgrounds 
$(b_n)=(t_1,0,t_3,0,0,\cdots)$ (purely NS-NS) 
and $(b_n)=(t_1,0,0,t_4,\cdots)$ (in a flow generated by R-R operator 
$\cO^{[1]}_4$) 
are given as follows 
(note that $\del=g\,\del/\del x^{[0]}_1=g\,\del/\del t_1$): 

\noindent
\underline{$\hat p=1,\hat q=2$}:
\begin{align}
\left\{
\begin{array}{ll}
0=3t_3 (H_-\del H_+-H_+\del H_-)+g\,\nu, \\
0=-2t_1H_+-3t_3 \bigl(\frac{1}{2}\del^2H_++4H_+^2H_-\bigr), \\
0=-2t_1H_--3t_3 \bigl(\frac{1}{2}\del^2H_-+4H_-^2H_+\bigr),
\end{array}
\right.
\label{str_eq_12}
\end{align}

\noindent
\underline{$\hat p=1,\hat q=3$}:
\begin{align}
\left\{
\begin{array}{ll}
0=4t_4(\frac{1}{2}H_+\del^2H_-+\frac{1}{2}H_-\del^2H_+ 
-\frac{1}{2}\del H_+\del H_- +6H_+^2H_-^2) + g\,\nu,\\
0=-2t_1H_+-4t_4 \bigl(\frac{1}{4}\del^3H_++6H_+H_-\del H_+\bigr), \\
0=-2t_1H_-+4t_4 \bigl(\frac{1}{4}\del^3H_-+6H_+H_-\del H_-\bigr),
\end{array}
\right.
\label{str_eq_13}
\end{align}
where the equations are displayed 
in the order of the diagonal, (1,2) and (2,1) elements.
Note that the diagonal equation including the flux $\nu$ 
is related with the other equations as
\begin{align}
\del ({\rm diagonal})=2\times \bigl[ H_-\times (1,2) - H_+\times (2,1)\bigr],
\end{align}
from which $\nu$ disappears. 
In fact, R-R flux $\nu$ can easily get lost 
in the analysis other than the Virasoro constraints. 
For example, instead of using the Virasoro constraints, 
we could obtain string equations 
by directly solving the Douglas equation $[\bP,\bQ]=g\bunit_2$. 
This calculation is performed in Appendix \ref{ZS}, 
and we find there that the obtained string equations 
are in the form of the commutator of eq.\ \eq{string_eq_p1} 
with $\sigma_3$ (see eq.\ \eq{StrEq-p1}), 
so that the information on $\nu$ is totally lost  
because $\nu$ appears in the diagonal elements of the Virasoro constraints. 
For the latter string equations \eq{StrEq-p1}, 
the diagonal element is a total derivative 
and $\nu$ is introduced as an integration constant \cite{HMPN,uni2dqg,ogu,UniCom,SeSh2}.  
This is a feature which holds generically for arbitrary $\hat p$, 
and shows the advantage of our string field theoretical approach; 
the R-R background flux $\nu$ has a definite meaning 
as the discrepancy between two Fermi levels 
and enters the expression without ambiguity.

\subsection{$\bZ_2$ symmetries and orbifolding}\label{CCt}

One can introduce the following $Z_2$ transformations 
into the theory:
\begin{align}
 {\rm C}:&~(\bsigma,\bL) \to (-\bsigma, \bL), \\
 \tilde{\rm C}:&~(\bsigma,\bL) \to (-\bsigma, -\bL). 
\end{align} 
Since the oscillator $\alpha^{[\mu]}_n=
\alpha^{(1)}_n+(-1)^\mu\alpha^{(2)}_n$ $(\mu=0,1)$ 
corresponds to the differential operator $(\bsigma^\mu \bL^n)_+$, 
the above transformations are rewritten in terms of oscillators as%
\footnote{
The charge conjugation in terms of infinite Grassmannian 
is described in Appendix \ref{C_Grassmannian}.
} 
\begin{align}
 {\rm C}:&~\alpha^{[\mu]}_n \to (-1)^\mu\,\alpha^{[\mu]}_n, 
  \qquad \hat \nu \to -\hat \nu \\
 \tilde{\rm C}:&~\alpha^{[\mu]}_n \to (-1)^{n+\mu}\,\alpha^{[\mu]}_n, 
  \quad \hat \nu \to -\hat \nu 
\end{align}
or
\begin{align}
 {\rm C}:&~\alpha^{(1)}_n \leftrightarrow \alpha^{(2)}_n, \\
 \tilde{\rm C}:&~\alpha^{(1)}_n \leftrightarrow 
   (-1)^n\,\alpha^{(2)}_n. 
\end{align}
Looking at the definition \eq{Winf_def} of the $W_{1+\infty}$ currents 
and noting that the generators have the following form in the twisted vacuum:
\begin{align}
 W^s_n=\sum_{r=0}^s\sum_{m_1+\cdots+m_r=n\hat p}\,
  w_{m_1\cdots m_r}\,\Bigl(\nol\alpha^{(1)}_{m_1}
  \cdots\alpha^{(1)}_{m_r}\nor+(-1)^{n}\nol\alpha^{(2)}_{m_1}
  \cdots\alpha^{(2)}_{m_r}\nor\Bigr),
\end{align} 
we find that under these $\bZ_2$ transformations 
the generators simply transform with multiplicative factors: 
\begin{align}
 {\rm C}:&~W^s_n \to (-1)^n\,W^s_n, \\
 \tilde{\rm C}:&~W^s_n \to (-1)^{n(\hat p+1)}\,W^s_n.
\end{align}
Thus, if a state $\ket\Phi$ satisfies the $W_{1+\infty}$ constraints, 
so does the transformed state $\ket{\Phi'}
={\rm C}\ket\Phi$ or $\tilde{\rm C}\ket\Phi$. 
Assuming the perturbative uniqueness of solutions 
to the $W_{1+\infty}$ constraints, 
we conclude that $\ket\Phi$ and $\ket{\Phi'}$ can differ 
only by a multiplication of D-instanton operators 
and we have the following perturbative identity for any operator $\cO$:
\begin{align}
 \frac{ \bra{b/g;\nu} \cO  \ket{\Phi} }{\bra{b/g;\nu}\Phi\bigr>}
  \,\biggl(= \frac{ \bra{{b'}/g;\nu'} \cO'  \ket{\Phi'} }
    {\bra{b'/g;\nu'} \Phi'\bigr>} \biggr)\,
  \simeq \frac{ \bra{{b'}/g;\nu'} \cO'  \ket{\Phi} }
    {\bra{b'/g;\nu'}\Phi\bigr>}.
 \label{perturbative_equivalence}
\end{align}
Here $b'=\bigl(b^{\prime\,(i)}_n\bigr)$ and $\cO'$ are 
the transforms of $b=\bigl(b^{(i)}_n\bigr)$ and $\cO$, respectively, 
under the transformation C or $\tilde{\rm C}$. 
In particular, if the backgrounds are invariant 
under one of such transformations, 
then the expectation values vanish perturbatively 
for those operators that are not invariant 
under the transformation.
Note that the purely NS-NS backgrounds where $b^{(1)}_n=b^{(2)}_n$ (or $b^{[1]}_n=0$) 
are invariant under the charge conjugation C, 
so that the correlation functions vanish perturbatively
if they contain odd number of R-R operators.

We introduce the orbifolding with the $\bZ_2$ transformation 
${\rm C}$ ({\em resp}.\ $\tilde{\rm C}$)
as the following condition on a decomposable fermion state 
satisfying the $W_{1+\infty}$ constraints: 
\begin{align}
 \alpha_n^{[1]}\ket{\Phi}=0 \ \ (\mbox{for}~{\rm C}),
  \qquad  \alpha_n^{[n+1]}\ket{\Phi}=0 \ \ (\mbox{for}~{\rm \tilde{C}}),
\end{align}
which is preserved along the 2cKP flows 
generated by $\alpha^{[0]}_n$  
({\em resp}.\  $\alpha^{[n]}_n$) alone. 
Under this condition, 
all the correlation functions including $\alpha_n^{[1]}$\ 
({\em resp}.\ $\alpha_n^{[n+1]}$) vanish, 
and thus these operators drop out of the spectrum. 
In particular,  type 0B theory with the C orbifolding 
is identified with type 0A theory, 
while theories of ${\hat p,\,\hat q \in 2\bZ+1}$ 
with the ${\rm \tilde C}$ orbifolding are identified with 
$(p,q)=(\hat p,\hat q)$ minimal superstrings. 

\section{$(p,q)$ minimal superstring field theory}

In this section, we construct field theory of $(p,q)$ minimal superstrings. 

\subsection{String fields and $\eta =-1$ FZZT branes}

According to our ansatz on operator identification given in subsection 
\ref{operator_identification}, 
the excitations in the NS-NS and R-R sectors 
are collected into the NS-NS and R-R scalars given by
\begin{align}
 \mbox{\underline{NS-NS}}:&\quad
  \del\varphi^{[0]}_0(\zeta)
  =\del\varphi^{(1)}_0(\zeta)+\del\varphi^{(2)}_0(\zeta)
  =\frac{1}{\hat p}\,\sum_{n\in\bZ}\,\alpha^{[0]}_n\,\zeta^{-n/\hat p-1},\\
 \mbox{\underline{R-R}}:&\quad
  \del\varphi^{[1]}_0(\zeta)
  =\del\varphi^{(1)}_0(\zeta)-\del\varphi^{(2)}_0(\zeta)
  =\frac{1}{\hat p}\,\sum_{n\in\bZ}\,\alpha^{[1]}_n\,\zeta^{-n/\hat p-1}.
\end{align}
Their connected correlation functions (or cumulants) in the presence 
of R-R flux $\nu$ are given by
\begin{align}
 \vev{\del\varphi_0^{(i_1)}(\zeta_1)\cdots
  \del\varphi_0^{(i_N)}(\zeta_N)}_{\!\nu,{\rm c}}
 =\biggl[\frac{\bra{b/g;\nu} :\!\del\varphi^{(i_1)}_0(\zeta_1)\cdots
   \del\varphi^{(i_N)}_0(\zeta_N)\!:\ket{\Phi}}
   {\bra{b/g;\nu}\,\Phi\bigr>}\biggr]_{\rm c}
\end{align}
and have an expansion in the string coupling $g$ as
\begin{align}
 \hspace{55mm}=\sum_{h\geq0}\,g^{2h+N-2}\,
  \vev{\,\del\varphi^{(i_1)}_0(\zeta_1)\cdots
  \del\varphi^{(i_N)}_0(\zeta_N)\,}^{\!(h)}_{\!\nu,{\rm c}}.
 \label{g_expansion2}
\end{align}
As in the bosonic case \cite{fy1}, 
the disk ($h=0,\,N=1$) and annulus ($h=0,\,N=2$) amplitudes 
have irregular terms 
which cannot be expressed with the local operators $\cO^{(i)}_n$:
\begin{align}
 \vev{\cO^{(i)}(\zeta)}_{\!\nu}
  &=\frac{1}{\hat p}\sum_{n=1}^\infty\,\vev{\cO^{(i)}_n}_{\!\nu}
   \,\zeta^{-n/\hat p-1}
  +g^{-1}\,N_1^{(i)}(\zeta) \label{1-point_expansion} \\
 \vev{\cO^{(i_1)}(\zeta_1)\,\cO^{(i_2)}(\zeta_2)}_{\!\nu,{\rm c}}
  &=\frac{1}{\hat p^2}\sum_{n_1,n_2=1}^\infty\,
  \vev{\cO^{(i_1)}_{n_1}\,\cO^{(i_2)}_{n_2}}_{\!\nu,{\rm c}}\,
  \zeta_1^{-n_1/\hat p-1}\zeta_2^{-n_2/\hat p-1}
  +g^{0}\,N_2^{(i_1 i_2)}(\zeta_1,\zeta_2)\label{2-point_expansion} \\
 \vev{\cO^{(i_1)}(\zeta_1)\cdots\cO^{(i_N)}(\zeta_N)}_{\!\nu,{\rm c}}
  &=\frac{1}{\hat p^N}\sum_{n_1,\cdots n_N=1}^\infty\,
  \vev{\cO^{(i_1)}_{n_1}\cdots\cO^{(i_N)}_{n_N}}_{\!\nu,{\rm c}}\,
  \zeta_1^{-n_1/\hat p-1}\cdots\zeta_N^{-n_N/\hat p-1}
  \quad (N\geq 3).
\end{align}
These terms (sometimes called ``nonuniversal terms'' though 
they are actually universal) are calculated in matrix models 
and found to be
\begin{align}
 N_1^{(i)}(\zeta)&=\frac{1}{\hat p}\,\sum_{n=1}^{\hat q+\hat p}
   n \,b^{(i)}_n\,\zeta^{n/\hat p-1}\\
 N_2^{(i_1 i_2)}(\zeta_1,\zeta_2)
  &=\delta^{i_1 i_2}\,\frac{\del}{\del\zeta_1}\frac{\del}{\del\zeta_2}
  \bigl[\ln\bigl(\lambda_1-\lambda_2\bigr)
   -\ln\bigl(\zeta_1-\zeta_2\bigr)\bigr]\nn\\
 &=\delta^{i_1 i_2}\,\Bigl[\,
\frac{d\lambda_1}{d\zeta_1}\frac{d\lambda_2}{d\zeta_2}
  \frac{1}{(\lambda_1-\lambda_2)^2}
   -\frac{1}{(\zeta_1-\zeta_2)^2}\Bigr]
 \qquad(\lambda=\zeta^{1/\hat p}).
\end{align}
Note that they do not depend on $\nu$.

By looking at the algebraic curves defined by disk amplitudes, 
we will see in the next section 
that $\varphi^{(i)}(\zeta)$ $(i=1,2)$ 
correspond to the boundary states of $\eta=-1$ charged FZZT branes: 
\begin{align}
\mbox{\underline{boundary states}}:\qquad
\ket{{\rm FZZT}+;\zeta}= \varphi_0^{(1)}(\zeta), \quad
\ket{{\rm FZZT}-;-\zeta}= \varphi_0^{(2)}(-\zeta).
\end{align}

Once a charged FZZT brane is located at a point 
in spacetime with coordinate $\zeta^2$, 
it becomes a source of fundamental strings, 
with a bunch of worldsheets which are not connected with each other 
in the sense of worldsheet topology, 
but are connected in spacetime 
with their boundaries pinched at the same superspace point $\zeta$. 
These configurations are easily summed up to give an exponential form 
as in \cite{fy3}, 
realizing the spacetime combinatorics of Polchinski \cite{combi}: 
\begin{align}
 \mbox{\underline{charged FZZT branes}}:\quad
  c^{(1)}_a(\zeta)=\,:\!e^{\varphi^{(1)}_a(\zeta)}\!:,\quad
  c^{(2)}_a(-\zeta)=\,:\!e^{\varphi^{(2)}_a(-\zeta)}\!:\quad
  (a=0,1,\cdots,\hat p-1).
 \label{charged_FZZT}
\end{align}
With these operators, the R-R charge operator 
$\hat \nu=(1/2)\,\bigl(\alpha^{(1)}_0-\alpha^{(2)}_0\bigr)
=(1/2\hat p)\,\sum_a\,\bigl(\alpha^{(1)}_{a,0}-\alpha^{(2)}_{a,0}\bigr)$ 
have the commutation relations
\begin{align}
  \bigl[\,\hat\nu, c^{(1)}_a(\zeta)\bigr]=\frac{1}{2}\,c^{(1)}_a(\zeta), \quad
   \bigl[\,\hat\nu, c^{(2)}_a(-\zeta)\bigr]=-\frac{1}{2}\,c^{(2)}_a(-\zeta), 
\end{align}
and thus, the generic partition functions in the presence of R-R flux $\nu$ 
are given by
\begin{align}
 B_{\nu}\bigl(\zeta^{(1)}_1,\cdots,\zeta^{(1)}_{N_1};
  \zeta^{(2)}_1,\cdots,\zeta^{(2)}_{N_2}\bigr) 
 \equiv 
  \dfrac{\Bigl<b/g;\,\nu+\frac{N_1-N_2}{2}\Bigr|\,
  \ds:\! \prod_{k=1}^{N_1}c^{(1)}_{a_k}(\zeta^{(1)}_k)\cdot
  \prod_{l=1}^{N_2}c^{(2)}_{b_l}(-\zeta^{(2)}_l)\!:\,
  \Bigl|\Phi\Bigr>}
  {\bigl<b/g;\,\nu\bigr|\Phi\bigr>}
\end{align}
with 
\begin{align}
 \bra{b/g;\,\nu+(N_1-N_2)/2}\equiv 
  \bra{b/g;\,\nu}\,e^{-N_1\bar\phi^{(1)}-N_2\bar\phi^{(2)}}.
\end{align}

All the operators introduced above have bosonized forms, 
and thus their correlation functions can be calculated (at least perturbatively) 
once the correlation functions 
$\bigl<\cO^{(i_1)}(\zeta_1)$ $\cdots$ 
$\cO^{(i_N)}(\zeta_N) \bigr>_{\!\nu,{\rm c}}$ 
are obtained.

\subsection{D-instanton operators and $\eta=-1$ ZZ branes}

As in the bosonic case \cite{fy1}, 
one can easily show that the commutator of the operator 
$c^{(i)}_a(\zeta^{(i)})\,\bar c^{(j)}_b(\zeta^{(j)})$ 
$(\zeta^{(1)}\equiv\zeta,\,\zeta^{(2)}\equiv-\zeta)$ 
with the generators of the $W_{1+\infty}$ algebra 
gives total derivatives 
unless $i=j$ and $a=b$, 
and thus the D-instanton operators \cite{fy1} 
\begin{align}
 D_{ab}^{(ij)}&=\epsilon_{ji}\,\gint\frac{d\zeta}{2\pi i}\,
  c_a^{(i)}(\zeta^{(i)})\bar c_b^{(j)}(\zeta^{(j)})
  =\gint\frac{d\zeta}{2\pi i}\,
  :\!e^{\varphi^{(i)}_a(\zeta^{(i)})-\varphi^{(j)}_b(\zeta^{(j)})}\!:
  \nn\\
 &(\mbox{$i=j$ with $a\neq b$; $i\neq j$ with $\forall(a,b)$})
\end{align}
commute with the generators:
\begin{align}
 \bigl[D_{ab}^{(ij)},\,W^s_n\bigr]=0.
\end{align}
This implies that for a given solution $\ket{\Phi}$ 
to the $W_{1+\infty}$ constraints, 
any product of D-instanton operators 
is again a solution. 
In order to keep the property that the resulting state is decomposable 
(i.e.\ can be written as $g\ket{0}$ 
with $g$ an exponential of fermion bilinears), 
the only possible form of such state is given by 
\begin{align}
 \ket{\Phi;\theta}\equiv 
  \prod_{i,j}\prod_{a,b}\,e^{\theta^{(ij)}_{ab}\,D^{(ij)}_{ab}}\ket\Phi
\end{align}
with constants $\theta^{(ij)}_{ab}$ \cite{fy2}. 
Note that we need to omit the case $i=j$ and $a=b$. 
Since $D^{(ij)}_{ab}$ includes the operator 
$e^{\bar\phi^{(i)}-\bar\phi^{(j)}}$,  
it changes the fermion number $(\nu^{(1)},\nu^{(2)})$ when $i\neq j$. 
We call $D^{(11)}_{ab}$ and $D^{(22)}_{ab}$ the neutral D-instanton operators 
and $D^{(12)}_{ab}$ and $D^{(21)}_{ab}$ the charged D-instanton operators.

We here write down their explicit form: 

\noindent
\underline{neutral D-instanton operators $(a\neq b)$}:
\begin{align}
 D_{ab}^{(11)}&=\gint\frac{d\zeta}{2\pi i}\,
  c_a^{(1)}(\zeta)\,\bar c_b^{(1)}(\zeta)
  =\gint\frac{d\zeta}{2\pi i}\,
  :\!e^{\varphi^{(1)}_a(\zeta)-\varphi^{(1)}_b(\zeta)}\!:,\\
 D_{ab}^{(22)}&=\gint\frac{d\zeta}{2\pi i}\,
  c_a^{(2)}(-\zeta)\,\bar c_b^{(2)}(-\zeta)
  =\gint\frac{d\zeta}{2\pi i}\,
  :\!e^{\varphi^{(2)}_a(-\zeta)-\varphi^{(2)}_b(-\zeta)}\!:.
\end{align}

\noindent
\underline{charged D-instanton operators $(\forall a,\forall b)$}:
\begin{align}
 D_{ab}^{(12)}&=-\gint\frac{d\zeta}{2\pi i}\,
  c_a^{(1)}(\zeta)\,\bar c_b^{(2)}(-\zeta)
  =\gint\frac{d\zeta}{2\pi i}\,
  :\!e^{\varphi^{(1)}_a(\zeta)-\varphi^{(2)}_b(-\zeta)}\!:,\\
 D_{ab}^{(21)}&=\gint\frac{d\zeta}{2\pi i}\,
  c_a^{(2)}(-\zeta)\,\bar c_b^{(1)}(\zeta)
  =\gint\frac{d\zeta}{2\pi i}\,
  :\!e^{\varphi^{(2)}_a(-\zeta)-\varphi^{(1)}_b(\zeta)}\!:.
\end{align}

When we insert a D-instanton operator $D_{ab}^{(ij)}$ 
into correlation functions 
and take the weak coupling limit $g\to+0$, 
the operator behaves as 
\begin{align}
 \langle D_{ab}^{(ij)}\rangle
  =\gint\frac{d\zeta}{2\pi i}\,
  e^{(1/g)\,\Gamma^{(ij)}_{ab}(\zeta) + \,O(g^0)}
 \label{Dinst_int}
\end{align}
with%
\footnote{
We see in the next section that the disk amplitudes do not depend 
on the R-R flux $\nu$.
} 
\begin{align}
 \Gamma^{(ij)}_{ab}(\zeta)
  =\bigl<\varphi^{(i)}_a(\zeta^{(i)})\bigr>^{(h=0)}
   -\bigl<\varphi^{(j)}_b(\zeta^{(j)})\bigr>^{(h=0)}.
\end{align}
If the integral has a saddle point $\zeta=\zeta^*$ 
and the contour can be deformed in such a way 
that it passes the saddle point in the standard direction 
and also that $\Gamma^{(ij)}_{ab}(\zeta)$ takes negative values 
all along the contour, 
then the presence of the D-instanton operator 
gives a controllable, finite nonperturbative effect, 
which can be evaluated in the vicinity of the saddle point \cite{fy1}. 
If we take a conformal background (defined later), 
the $\Gamma^{(ij)}_{ab}(\zeta)$ at saddle points 
give the partition functions of stable ZZ branes \cite{FIS,fim}.

\section{Algebraic curves for $\eta=-1$ FZZT branes}

The algebraic curves defined by the FZZT branes 
$\varphi^{(i)}_0(\zeta)$ $(i=1,2)$ can be calculated 
from the $W_{1+\infty}$ constraints with proper boundary conditions 
supplied by the structure of 2cKP hierarchy. 
This procedure is totally the same with the one for the bosonic case 
\cite{fim}.

We first note that the disk amplitudes 
\begin{align}
 Q^{(i)}(\zeta)&\equiv\,\lim_{g\to 0}\,g\,
  \frac{
   \bra{b/g;\,\nu}\,\del\varphi^{(i)}_0(\zeta)\,\ket{\Phi}
  }{
  \bra{b/g;\,\nu}\,\Phi\bigr>
  } \nn\\
 &=
  \frac{1}{\hat p}\,\sum_{n=1}^{\hat q+\hat p}\,n\,b^{(i)}_n\,\zeta^{n/\hat p-1}
  + \frac{1}{\hat p}\,\sum_{n=1}^\infty\,v^{(i)}_n\,\zeta^{-n/\hat p-1}, 
 \label{disk}
\end{align}
do not depend on $\nu$ when $\nu^{(1)}=-\nu^{(2)}=\nu$ is finite. 
Here $v^{(i)}_n\equiv\lim_{g\to 0}\,g\,\bigl<\alpha^{(i)}_n\bigr>$ 
is the expectation value of the operator $\alpha^{(i)}_n$ on sphere. 
In fact, as for the first term 
it should be clear that it does not depend on $\nu$. 
As for the second term, the $v^{(i)}_n$ is the expectation value 
taken for the background $\bra{b/g;\,\nu}$, 
and thus has a potential dependence on $\nu$. 
However, the expectation values taken for $\bra{b/g;\,\nu}$ 
and for $\bra{b/g;\,\nu=0}\equiv\bra{b/g}$ 
differ only by a next leading term, 
so that we can safely neglect the presence of $\nu$ 
in calculating disk amplitudes.  
We comment that if the R-R flux $\nu^{(i)}$ is tuned as 
$\nu^{(i)}(g)=\tilde\nu^{(i)}/g$, 
then an extra term $\tilde\nu^{(i)}/\hat p\,\zeta$ 
appears in the last expression of \eq{disk}.

The master field equation to obtain disk amplitudes is given by 
the fact that the expectation values of the $W_{1+\infty}$ currents 
are polynomials in $\zeta$ (see \eq{Winf_zeta}):
\begin{align}
 \bigl< W^s(\zeta) \bigr>_\nu
  \,\,\biggl(=\frac{
  \bigl<b/g;\,\nu\bigr|\,W^s(\zeta)\,\bigl|\Phi\bigr>
  }{
  \bigl<b/g;\,\nu\bigr|\Phi\bigr>
  } \biggr)\,
  \equiv s\,a_s(\zeta)/g^s.
\end{align}
Decomposing the $W_{1+\infty}$ currents as a sum over $a=0,1,\cdots,\hat p-1$ 
as in \eq{Winf_def}, 
the above equation is rewritten as (see \cite{fim})%
\footnote{
Here for a series expansion of a function $f(\zeta)=\sum_{n\in\bZ}
f_n\,\zeta^{n/\hat p}$, 
we define its polynomial part as 
$\bigl[f(\zeta)\bigr]_{\rm pol}\equiv\sum_{m\geq 0}f_{m\hat p}\,\zeta^m$.
} 
\begin{align}
 \sum_{a=0}^{\hat p-1}\,\bigl< \cW^s_a(\zeta)\bigr>_\nu
  =\hat p\,\bigl[\,\bigl< \cW^s_0(\zeta)\bigr>_\nu\bigr]_{\rm pol}
  = s\,a_s(\zeta)/g^s.
\end{align}
In the weak coupling limit, 
$\vev{\cW^a_s(\zeta)}$ receives the dominant contributions 
from the products of one-point functions with maximal number 
(equal to $s$), 
and thus we have 
\begin{align}
 \lim_{g\to0}g^s\vev{\cW^a_s(\zeta)}
  =\sum_{a=0}^{\hat p-1}\Bigl[
  \bigl(Q^{(1)}_a(\zeta)\bigr)^s
  +\bigl(-Q^{(2)}_a(-\zeta)\bigr)^s \Bigr]=s\,a_s(\zeta).
 \label{Miwa}
\end{align}
Here we have introduced
\begin{align}
 Q^{(i)}_a(\zeta)\equiv \vev{\del\varphi_a^{(i)}(\zeta)}^{(0)}\quad
  (i=1,2).
\end{align}
The algebraic curve is then defined by
\begin{align}
 F(\zeta,Q)\equiv 
  \prod_{a=0}^{\hat p-1}\bigl(Q-Q_a^{(1)}(\zeta)\bigr)
  \cdot \prod_{a=0}^{\hat p-1}\bigl(Q+Q_a^{(2)}(-\zeta)\bigr)=0.
 \label{curve}
\end{align}
This is actually a polynomial both of $\zeta$ and $Q$. 
In fact, if we set $2\hat p$ variables as
\begin{align}
 Q_r(\zeta)\equiv
 \begin{cases}
  Q^{(1)}_a(\zeta) & (r=a=0,1,\cdots,\hat p-1), \\
  -Q^{(2)}_a(-\zeta) & (r=\hat p+a=\hat p,\,\hat p+1,\cdots,2\hat p-1),
 \end{cases}
\end{align}
then $F(\zeta,Q)$ is expanded in $Q$ with the coefficients 
which are polynomials of the variables $a_s(\zeta)$:
\begin{align}
 F(\zeta,Q)&=\prod_{r=0}^{2\hat p-1}\,\bigl(Q-Q_r(\zeta)\bigr)
  =Q^{2\hat p}\cdot \exp\Bigl[\sum_{r=0}^{2\hat p-1}
   \ln\Bigl(1-\frac{Q_r(\zeta)}{Q(\zeta)}\Bigr)\Bigr]\nn\\ 
 &=Q^{2\hat p}\cdot \exp\Bigl[-\sum_{s\geq 1}\,a_s(\zeta)\,Q^{-s}\Bigr]\nn\\
 &=\sum_{k=0}^{2\hat p}\,\cS_k\bigl([-a_s(\zeta)]\bigr)\,Q^{2\hat p-s}.
\end{align}

The curve \eq{curve} has a solution $Q=Q_0^{(1)}(\zeta)$, 
and thus describes the FZZT branes carrying positive charge. 
The curve for negative charge must have a solution $Q=Q^{(2)}(\zeta)$ 
and thus is given by the curve
\begin{align}
 F^{\rm C}(\zeta,Q)
  =\prod_{a=0}^{\hat p-1}(Q-Q_a^{(2)}(\zeta))
  \cdot \prod_{a=0}^{\hat p-1}(Q+Q_a^{(1)}(-\zeta))
  =F(-\zeta,-Q). 
\end{align}
Note that the curve has a dependence on the backgrounds $b=(b_n)$, 
$F(\zeta,Q;b)=0$, 
and the curve for negative charge branes 
are obtained by replacing the backgrounds 
in the curve for positive charge branes with their charge conjugates:
\begin{align}
 F^{\rm C}(\zeta,Q;b)
  \,\Bigl(=F(-\zeta,-Q;b)\Bigr)
  =F(\zeta,Q;b^{\rm C}).
 \label{curve^c}
\end{align}
In fact, using in the perturbative equality \eq{perturbative_equivalence} 
the fact that the charge conjugation exchanges 
$c_a^{(1)}(\zeta)$ and $c_a^{(2)}(\zeta)$ 
and that $\nu$ can be neglected for one-point functions $Q^{(i)}(\zeta)$, 
we have 
\begin{align}
 Q^{(1)}_a(\zeta;b)
 &=\lim_{g\to0}\,g\,\frac{
  \bra{b/g}\,\del\varphi^{(1)}_a(\zeta)\,\ket{\Phi}
  }{
  \bra{b/g}\Phi\bigr>
  }
 =\lim_{g\to0}\,g\,\frac{
  \bra{b^{\rm C}/g}\,\del\varphi^{(2)}_a(\zeta)\,
  \ket{\Phi^{\rm C}}
  }{
  \bra{b^{\rm C}/g}\Phi^{\rm C}\bigr>
  }\nn\\
 &=\lim_{g\to0}\,g\,\frac{
  \bra{b^{\rm C}/g}\,\del\varphi^{(2)}_a(\zeta)\,
  \ket{\Phi}
  }{
  \bra{b^{\rm C}/g}\Phi\bigr>
  }
 = Q^{(2)}_a(\zeta;b^{\rm C}), 
\end{align}
so that we obtain the relation, 
$a_s(\zeta;b)=(-1)^s\,a_s(-\zeta;b^{\rm C})$, 
which proves \eq{curve^c}.

\subsection{1-cut and 2-cut solutions}\label{1-cut&2-cut}

Generically the functions $a_s(\zeta)$ 
become polynomials for all $s\geq1$ 
only when we add the two terms in \eq{Miwa}, 
giving general 2-cut solutions. 
However, for some particular states $\ket{\Phi}$ 
it happens that each of the two gives a polynomial separately:
\begin{align}
 \sum_{a=0}^{\hat p-1} \bigl(Q^{(1)}_a(\zeta)\bigr)^s = s\,a^{(1)}_s(\zeta),
  \quad
  \sum_{a=0}^{\hat p-1}
  \bigl(Q^{(2)}_a(\zeta)\bigr)^s =s\,a^{(2)}_s(\zeta) 
  \quad(\forall s\geq 1).
 \label{1-cut}
\end{align}
This happens in the case when the state is factorized as
\begin{align}
 \ket{\Phi}=\ket{\Phi^{(1)}}\otimes \ket{\Phi^{(2)}}
  = g^{(1)}\otimes g^{(2)}\,\ket{0}
 \label{1-cut2}
\end{align}
with $\ket{\Phi^{(i)}}$ constructed only with $\phi^{(i)}(\lambda)$ 
and satisfying the $W_{1+\infty}$ constraints 
independently:
\begin{align}
 (W^{(1)})^{s}_n\ket{\Phi}=(W^{(2)})^{s}_n\ket{\Phi}=0 
 \qquad (s\geq1; \ n\ge -s+1),
\end{align}
where $(W^{(i)})^s(\zeta)=\sum_n (W^{(i)})^s_n\,\zeta^{-n-s} 
\equiv\sum_{a=0}^{\hat p-1}:\!
e^{-\varphi_a^{(i)}(\zeta)}\del^se^{\varphi_a^{(i)}(\zeta)}
\!:$. 
In this case, the algebraic equation is given by the product 
of polynomials: 
\begin{align}
 F(\zeta,Q)=(-1)^{\hat p}\,F^{(1)}(\zeta,Q)\cdot F^{(2)}(-\zeta,-Q)=0.
 \label{1-cut_alg}
\end{align}
The solution is thus expressed by the algebraic curve $F^{(1)}(\zeta,Q)=0$, 
and we call this a 1-cut solution. 
Actually, in order to realize the situation \eq{1-cut} alone, 
we can multiply the state \eq{1-cut2} with D-instanton operators 
since their nonperturbative contributions do not appear 
in the algebraic curve. 
Since neutral D-instanton operators can be absorbed 
into $\ket{\Phi^{(1)}}$ or $\ket{\Phi^{(2)}}$, 
we only need to consider the charged D-instanton operators. 
We thus obtain the general form of a state corresponding to 
1-cut solutions:
\begin{align}
 \ket{\Phi}
  =\prod_{a,b}\,e^{\theta^{(12)}_{ab}\,D^{(12)}_{ab}+\,
  \theta^{(21)}_{ab}\,D^{(21)}_{ab}}\,
  \ket{\Phi^{(1)}}\otimes \ket{\Phi^{(2)}}.
\end{align}
Note that once we take into account the nonperturbative effects 
from the charged D-instanton operators, 
the functions $a^{(1)}_s(\zeta)$ and $a^{(2)}_s(\zeta)$ 
are no longer polynomials of $\zeta$ separately, 
and the disk amplitudes are better described 
by the curves of 2-cut solutions.

We comment that only 1-cut solutions are allowed 
for such backgrounds that have both of C 
and $\tilde{\rm C}$ symmetries. 
Recall that we impose the $\tilde{\rm C}$ invariance 
only on odd minimal superstrings, 
$(p,q)=(\hat p,\hat q)$ 
with $\hat p,\,\hat q\in 2\bZ+1$. 
To prove the statement, it is sufficient to show the following equality 
(shown at the end of this subsection):
\begin{align}
 \sum_a\bigl(-Q^{(2)}_a(-\zeta)\bigr)^s= \sum_a\bigl(Q^{(1)}_a(\zeta)\bigr)^s,
 \label{odd_model}
\end{align}
which together with \eq{Miwa} implies that $a^{(1)}_s(\zeta)
=(-1)^s a^{(2)}_s(-\zeta)=(1/2)a_s(\zeta)$ 
(polynomial). 
Furthermore, the second factor in the algebraic equation \eq{1-cut_alg} 
can be rewritten as 
\begin{align}
 F^{(2)}(-\zeta,-Q)
  &=\sum_{k=0}^{\hat p-k}\,\cS_k([-a^{(2)}_s(-\zeta)])\,(-Q)^{\hat p-k}\nn\\
 &=\sum_{k=0}^{\hat p-k}\,\cS_k([-(-1)^s\,a^{(1)}_s(\zeta)])\,(-Q)^{\hat p-k}\nn\\
 &=(-1)^{\hat p}\,\sum_{k=0}^{\hat p-k}\,
  \cS_k([-\,a^{(1)}_s(\zeta)])\,Q^{\hat p-k}\nn\\
 &=(-1)^{\hat p}\,F^{(1)}(\zeta,Q),
\end{align}
so that the algebraic equation has the complete square form:
\begin{align}
 F(\zeta,Q) =\bigl(F^{(1)}(\zeta,Q)\bigr)^2=0\quad
 (\mbox{for C- and $\tilde{\rm C}$-invariant backgrounds}).
\end{align}

We conclude this subsection with showing eq.\ \eq{odd_model} 
for C- and $\tilde{\rm C}$-invariant backgrounds.
We first rewrite the left-hand side as
\begin{align}
 \sum_a\bigl(-Q^{(2)}_a(-\zeta)\bigr)^s
  &= \sum_a\bigl(-Q^{(1)}_a(-\zeta)\bigr)^s\quad(\mbox{C invariance})\nn\\
 &=\sum_a\bigl(-Q^{(1)}_0(-e^{2\pi ia}\zeta)\bigr)^s\quad 
    \Bigl( Q^{(1)}_0(\zeta)=(1/\hat p)\sum_{n\in\bZ}
    v^{(1)}_n\zeta^{-n/\hat p-1}\Bigr)\nn\\
 &=\frac{1}{\hat p^{s-1}}\,\sum_{n\in\bZ}\sum_{n_1\cdots+n_s=n\hat p}
  v^{(1)}_{n_1}\cdots v^{(1)}_{n_s}\,(-1)^n\,\zeta^{-n-s}.
\end{align}
Only the terms with $n\in 2\bZ$ survive in the last expression 
since $v^{(1)}_{n_r}=0$ for odd $n_r$ 
(due to the C$\times\tilde{\rm C}$ invariance), and $\hat p$ is odd now.
Thus we have 
\begin{align}
 =&\,\frac{1}{\hat p^{s-1}}\,\sum_{n\in\bZ}
  \sum_{n_1\cdots+n_s=n\hat p}
  v^{(1)}_{n_1}
  \cdots v^{(1)}_{n_s}\,\zeta^{-n-s}\nn\\
 =&\,\sum_a\bigl(Q^{(1)}_a(\zeta)\bigr)^s.
\end{align}

\subsection{Super Kazakov series $(\hat p=1)$ }

In this case, the disk amplitude is expressed as
\begin{align}
 Q^{(i)}(\zeta)\equiv \vev{\del\varphi^{(i)}(\zeta)}^{(h=0)}
  =\sum_{n=1}^{1+\hat q}n\,b_n^{(i)} \zeta^{n-1}
   +\sum_{n\geq1}v^{(i)}_n\,\zeta^{-n-1}.
\end{align}
The constraints $W^1_n\ket\phi=0$ $(n\geq 0)$ 
imply $v^{[n]}_n=0$, 
and the perturbations with respect to unphysical variables 
$b^{[n]}_n$ can always be absorbed into a shift of $Q^{(i)}$ 
as in the bosonic case \cite{fim}, 
\begin{align}
 Q^{(1)}(\zeta)&\to Q^{(1)}(\zeta)
  +\sum_{n=1}^{1+\hat q}n\,b^{[n]}_n\zeta^{n-1}
  =Q^{(1)}(\zeta)+\frac{1}{2}\,a_1(\zeta),\\
 Q^{(2)}(\zeta)&\to Q^{(2)}(\zeta)
  +\sum_{n=1}^{1+\hat q}n\,(-1)^n b^{[n]}_n\zeta^{n-1}
  =Q^{(2)}(\zeta)-\frac{1}{2}\,a_1(-\zeta),
\end{align}
so that one can set $a_1(\zeta)=0 \ (\Leftrightarrow b_n^{[n]}=0)$ without loss of generality.
Then it holds that $Q^{(1)}(\zeta)=Q^{(2)}(-\zeta)\equiv Q(\zeta)$,%
\footnote{
In the $\hat p =1$ case  with the condition $b_n^{[n]}=0$, 
the two-component system can be completely written 
by using a single untwisted boson 
$\del\phi(\lambda)=\del \varphi(\zeta)\equiv \del\varphi^{(1)}(\zeta)
=\del\varphi^{(2)}(-\zeta)$,  
as was pointed out in \cite{CDM}.
} 
and we have 
\begin{align}
 a_2(\zeta)=\frac{1}{2}\,\bigl[ \bigl(Q^{(1)}(\zeta)\bigr)^2
  +\bigl(-Q^{(2)}(-\zeta)\bigr)^2\bigr]_{\rm pol}
  =\bigl[ \bigl(Q(\zeta)\bigr)^2\bigr]_{\rm pol}. 
\end{align}
Thus, we have the curve
\begin{align}
 F(\zeta,Q)=Q^2-a_2(\zeta)=0.
\end{align}

\noindent
\underline{\bf 1-cut solutions:}

Due to the general consideration given in the previous subsection, 
1-cut solutions are given by the state $\ket\Phi$ satisfying 
the $W_{1+\infty}$ constraints separately (at least perturbatively), 
\begin{align}
 (W^{(1)})^s_n\ket\Phi=0,\quad (W^{(2)})^s_n\ket\Phi=0 \qquad 
(s\geq1; \ n\ge -s+1).
\end{align}
In particular, this implies that 
\begin{align}
 F^{(i)}(\zeta,Q)=Q-a^{(i)}_1(\zeta)=Q-\sum_{n=1}^{1+\hat q}
  n\,b_n^{(i)}\zeta^{n-1}=0 \quad(i=1,2),
\end{align}
and we have
\begin{align}
 Q(\zeta)=\sum_{n=1}^{1+\hat q}  n\,b_n^{(i)}\zeta^{n-1}
  \quad(\mbox{1-cut solutions}).
 \label{superK_1cut}
\end{align}
Note that all the one-point functions vanish, $v^{(i)}_n=0$.

\noindent
\underline{\bf 2-cut solutions:}

Based on the analysis of the $\hat p=1$ string equations, 
we take boundary conditions such that 2-cut solutions have 
cuts in the complex $\zeta$ plane, 
each of which has one of its end points at infinity.

The master field equation is now given by
\begin{align}
 a_2(\zeta)=\bigl[ \bigl(Q(\zeta)\bigr)^2\bigr]_{\rm pol}
  =\Bigl( \sum_{n=1}^{\hat q+1}n\,b_n\zeta^{n-1}\Bigr)^2+f(\zeta),
\end{align}
where $f(\zeta)$ is a polynomial of degree $\hat q-2$ 
depending on the one-point functions $v^{(i)}_n$,  
and has $\hat q-1$ degrees of freedom. 
They can be adjusted by properly choosing $v^{(i)}_n$ 
such that the equations
\begin{align}
 F(\zeta^*,Q^*)
 =\frac{\del F}{\del\zeta}(\zeta^*,Q^*)
 =\frac{\del F}{\del Q}(\zeta^*,Q^*)
 =0
\end{align}
have its maximal $\hat q-1$ solutions. 
This is equivalent to the condition 
that $a_2(\zeta)$ has $\hat q-1$ solutions 
to the equations
\begin{align}
 a_2(\zeta^*)=\frac{\del a_2}{\del \zeta}(\zeta^*)=0. 
\end{align}
The solutions are given by%
\footnote{
The general form of the irrational function 
would be $\sqrt{(\zeta-a)(\zeta-b)}$, 
but this can always be changed to  $\sqrt{\zeta^2+a^2}$ 
by shifting $\zeta$ properly.
} 
\begin{align}
 Q(\zeta)=E(\zeta)\cdot\sqrt{\zeta^2+a^2}
 \label{superK_2cut}
\end{align}
with a polynomial $E(\zeta)={\rm const.}
\prod_{r=1}^{\hat q-1}(\zeta-\zeta^*_r)$. 

\subsection{Conformal backgrounds}

\noindent
\underline{\bf 1-cut solutions:}

The conformal backgrounds for 1-cut solutions 
are given by C-invariant backgrounds $b^{(1)}_n=b^{(2)}_n=b_n$ with 
\begin{align}
 b_n &= 
  \begin{cases}\ds
   -\beta\,
    \frac{\hat p\,\hat q}{n}\,
    \frac{2^{(\hat q-\hat p)/\hat p}}{2m\hat p-\hat q}
    \binom{2m-\hat q/\hat p}{m}\Bigl(\frac{a^{\hat p}}{4}\Bigr)^{m}  
      & \biggl(n=\hat q+\hat p-2m\hat p\,;
     ~0\leq m\leq \Bigl[\dfrac{\hat q+\hat p-1}{2\hat p}\Bigr]\biggr) \\
    0 & \bigl(\hbox{otherwise}\bigr)
  \end{cases},\nn\\
&~~~~~~(\beta:~\hbox{numerical~constant}). 
 \label{conf_bgds_1-cut}
\end{align}
They can be solved in the same manner 
with that in the bosonic conformal backgrounds
(see eq.\ (3.47) of \cite{fim}); 
one simply needs to replace their $(p,q)$ by $(\hat p,\hat q)$. 
For this, the curve for $Q(\zeta)=Q^{(1)}(\zeta)$ is given by 
$F^{(1)}(\zeta,Q)=0$. 
As is shown in \cite{fim}, we have the uniformization parameter $z$ 
which parametrizes $(\zeta,Q)$ as%
\footnote{
This $z$ corresponds to the derivative $\del=\del/\del x^{[0]}_1$ 
in the weak coupling limit, with the relation $z=2^{1-1/\hat p}\,\del$.
} 
\begin{align}
 z&=a\,\cosh\tau,\\
 \zeta&=a^{\hat p}\,\cosh\hat p\tau,\\
 Q&=\beta\,a^{\hat q}\,\cosh\hat q\tau,
\end{align}
giving the curve 
\begin{align}
 F^{(1)}(\zeta,Q)=2^{1-\hat p}\,\beta^{\hat p}\,a^{\hat p\hat q}\,
 \Bigl[\,
 T_{\hat p}\bigl(Q/\beta a^{\hat q}\bigr)
 -T_{\hat q}\bigl(\zeta/a^{\hat p}\bigr)\Bigr]=0, \label{conalg1cut}
\end{align}
where $T_n(z)$ is the first Tchebycheff polynomials of degree $n$, 
$T_n(\cosh\tau)=\cosh(n\tau)$. 
Taking a branch such that $\zeta/z^{\hat p}>0$ for ${\rm Re}\,z\to\infty$, 
we have
\begin{align}
 Q_0(\zeta) = \frac{\beta}{2}\,
    \left[\bigl(\zeta+\sqrt{\zeta^2-a^{2\hat p}}\bigr)^{\hat q/\hat p}
   +\bigl(\zeta-\sqrt{\zeta^2-a^{2\hat p}}\bigr)^{\hat q/\hat p}\right].
 \label{conf_Q_1-cut}
\end{align}
This agrees with the curve of $\eta=-1$ FZZT branes 
in super Liouville theory \cite{SeSh}
if one identifies $a=\mu^{1/2\hat p}$ $(\mu>0)$.
One can easily see that eq.\ \eq{superK_1cut} is reproduced 
when $\hat p=1$.

Note that the total algebraic equation $F(\zeta,Q)=0$ is expressed as 
\begin{align}
F(\zeta,Q)=2^{1-2\hat p}\beta^{2\hat p}a^{2\hat p \hat q}
\Bigl[\,
 T_{2\hat p}\bigl(Q/\beta a^{\hat q}\bigr)
 -T_{2\hat q}\bigl(\zeta/a^{\hat p}\bigr)\Bigr]=0
\end{align}
for even models, and 
\begin{align}
F(\zeta,Q)=2^{2-2\hat p}\beta^{2\hat p}a^{2\hat p \hat q}
\Bigl[\,
 T_{\hat p}\bigl(Q/\beta a^{\hat q}\bigr)
 -T_{\hat q}\bigl(\zeta/a^{\hat p}\bigr)\Bigr]^2=0
\end{align}
for odd models. 
Thus, even and odd models have the same description for their curves. 
By requiring that the odds models 
are under the ${\rm \tilde C}$ orbifolding, 
the curves are reduced to \eq{conalg1cut}. 
This explains why odd models seemed different in Liouville theory \cite{SeSh}.

\noindent
\underline{\bf 2-cut solutions:}

The conformal backgrounds for 2-cut solutions 
are given by C-invariant backgrounds $b^{(1)}_n=b^{(2)}_n=b_n$ with 
\begin{align}
 b_n &= 
  \begin{cases}\ds
   -\beta\,
    \frac{\hat p\,\hat q}{n}\,
    \frac{2^{(\hat q-\hat p)/\hat p}}{2m\hat p-\hat q}
    \binom{2m-\hat q/\hat p}{m}\Bigl(-\frac{a^{2\hat p}}{4}\Bigr)^{m}  
      & \biggl(n=\hat q+\hat p-2m\hat p\,;
     ~0\leq m\leq \Bigl[\dfrac{\hat q+\hat p-1}{2\hat p}\Bigr]\biggr) \\
    0 & \bigl(\hbox{otherwise}\bigr)
  \end{cases}\nn\\
&~~~~~~(\beta:~\hbox{numerical~constant}). 
 \label{conf_bgds_2-cut}
\end{align}
Note that they could lead to C- and $\tilde{\rm C}$-invariant backgrounds 
if we took $\hat p+\hat q\in 2\bZ$. 
Thus, conformal backgrounds are allowed to have 2-cut solutions 
only for even minimal superstrings, 
$(p,q)=(2\hat p,2\hat q)$ with $\hat p+\hat q\in 2\bZ+1$, 
which we assume in the following discussions.

For this, the curve $F(\zeta,Q)=0$ for $Q(\zeta)=Q^{(1)}(\zeta)$ 
is not factorized, 
and is parametrized as
\begin{align}
 z&=a\,\cosh\tau,\\
 \zeta&=a^{\hat p}\,\sinh\hat p\tau,\\
 Q&=\beta\,a^{\hat q}\,\sinh\hat q\tau,
\end{align}
giving a curve
\begin{align}
 F(\zeta,Q)=(-1)^{\hat p}\,2^{1-2\hat p}\,\beta^{2\hat p}\,a^{2\hat p \hat q}\,
 \Bigl[\,T_{2\hat p}(i Q/\beta a^{\hat q})
  +T_{2\hat q}(i \zeta/a^{\hat p})\Bigr]
 =0\quad(\hat p+\hat q\in 2\bZ+1).
 \label{curve_2-cut}
\end{align}
Taking a branch such that $\zeta/z^{\hat p}>0$ for ${\rm Re}\,z\to\infty$, 
we have
\begin{align}
 Q_0(\zeta) = \frac{\beta}{2}\,
    \left[\bigl(\zeta+\sqrt{\zeta^2+a^{2\hat p}}\bigr)^{\hat q/\hat p}
   -\bigl(-\zeta+\sqrt{\zeta^2+a^{2\hat p}}\bigr)^{\hat q/\hat p}\right].
 \label{conf_Q_2-cut}
\end{align}
This agrees with the curve of $\eta=-1$ FZZT branes 
in super Liouville theory \cite{SeSh}
if one identifies $a=|\mu|^{1/2\hat p}$ $(\mu<0)$.
Note that eq.\ \eq{superK_2cut} is reproduced when $\hat p=1$.

\section{$\eta=-1$ ZZ branes}

In order to make calculations definite, 
we take the conformal backgrounds 
\eq{conf_bgds_1-cut} for 1-cut solutions 
and \eq{conf_bgds_2-cut} for 2-cut solutions. 
Then the saddle points in \eq{Dinst_int} describe ZZ branes \cite{fy1,FIS,fim}%
\footnote{A similar analysis could be performed in terms of matrix models by 
extending the analysis of \cite{KazakovKostov}}.
Recalling that the relations $Q^{(1)}_a(\zeta)=Q^{(2)}_a(\zeta)$ hold 
in the conformal backgrounds, 
we have
\begin{align}
 \mbox{\underline{neutral D-instantons:}}~~~
  \Gamma^{(11)}_{ab}(\zeta)
  &=\Gamma^{(22)}_{ab}(-\zeta)\nn\\
  &=\bigl<\varphi^{(1)}_a(\zeta)\bigr>^{(h=0)}
    -\bigl<\varphi^{(1)}_b(\zeta)\bigr>^{(h=0)}
  \quad(a\neq b),\\
 \mbox{\underline{charged D-instantons:}}~~~
  \Gamma^{(12)}_{ab}(\zeta)
  &=\Gamma^{(21)}_{ab}(-\zeta)\nn\\
  &=\bigl<\varphi^{(1)}_a(\zeta)\bigr>^{(h=0)}
    -\bigl<\varphi^{(1)}_b(-\zeta)\bigr>^{(h=0)}
  \quad(\forall a,\forall b).
\end{align}

\subsection{Neutral ZZ branes}

The $\Gamma^{(11)}_{ab}$ is expressed 
as a function of $\tau$ 
by integrating $d\Gamma^{(11)}_{ab}/d\tau
=(d\Gamma^{(11)}_{ab}/d\zeta)\cdot(d\zeta/d\tau)$. 
Noting that
\begin{align}
 \frac{d\Gamma^{(11)}_{ab}}{d\zeta}
  =Q^{(1)}_a\bigl(\zeta(\tau)\bigr)-Q^{(1)}_b\bigl(\zeta(\tau)\bigr) 
  =Q^{(1)}_0\bigl(\zeta(\tau_a)\bigr)-Q^{(1)}_0\bigl(\zeta(\tau_b)\bigr)
 \label{saddle_pt_eq_neutral}
\end{align} 
with $\tau_a\equiv \tau+2\pi ia/\hat p$,
we obtain
\begin{align}
 \Gamma_{ab}
 &=\beta\,\hat p\,a^{\hat p+\hat q}
  \biggl[\, 
  \frac{1}{\hat q+\hat p}\Bigl(\cosh(\hat p\tau+\hat q\tau_a\bigr)
   -\cosh\bigl(\hat p \tau +\hat q\tau_b\bigr)\Bigr) \mp \nn\\
 &~~~~~~~~~~~~\mp \frac{1}{\hat q-\hat p}\Bigl(\cosh(\hat p\tau-\hat q\tau_a\bigr)
   -\cosh\bigl(\hat p \tau-\hat q\tau_b\bigr)\Bigr)\biggr],
\end{align}
where the upper ({\em resp}.\ lower) sign corresponds
to 1-cut ({\em resp}.\ 2-cut) solutions.  
The saddle points are found by solving  \eq{saddle_pt_eq_neutral} 
as in \cite{FIS,fim}:%
\footnote{
Other saddle points corresponding to $d\zeta/d\tau=0$ do not give 
major contributions to the contour integral, since the Jacobian vanishes there 
\cite{FIS}.
} 
\begin{align}
 \tau^*=\begin{cases} 
  \dfrac{1}{2}\,\Bigl(-\dfrac{2(a+b)}{\hat p}+\dfrac{2k}{\hat q}\Bigr)\,
   \pi i & (\mbox{1-cut}) \\
  \dfrac{1}{2}\,\Bigl(-\dfrac{2(a+b)}{\hat p}+\dfrac{2k+1}{\hat q}\Bigr)\,
   \pi i \quad & (\mbox{2-cut})
 \end{cases}
 \quad (k\in\bZ).
 \label{ZZ_neutral1}
\end{align}
Setting $m=2(b-a)$ for both of 1-cut and 2-cut solutions, 
and $n=2k$ for 1-cut and $n=2k+1$ for 2-cut solutions, 
we have
\begin{align}
 \tau^*_a&=\frac{1}{2}\Bigl(-\frac{m}{\hat p}+\frac{n}{\hat q}\Bigr)\,\pi i,
 \label{ZZ_neutral2}\\
 \tau^*_b&=\frac{1}{2}\Bigl(+\frac{m}{\hat p}+\frac{n}{\hat q}\Bigr)\,\pi i.
 \label{ZZ_neutral3}
\end{align}
Taking into account the fact that distinct ZZ branes 
are labeled by the indices $(m,n)$ with 
$\hat q m-\hat p n\ge 0$ 
and also counting the contributions from $\Gamma_{ab}^{(22)}$, 
the number of neutral ZZ branes is given by 
\begin{align}
\begin{cases} 
  \dfrac{(\hat p-1)(\hat q-1)}{2}\times 2 & (\mbox{1-cut}) \\
  \Bigl[\dfrac{(\hat p-1)\hat q+1}{2}\Bigr]\times 2  & (\mbox{2-cut}),
 \end{cases}
\end{align}
where $[\cdots]$ denotes the integer part.%
\footnote{We also count an instanton with 
$\hat qm-\hat pn=0$, i.e.\ $(m,n)=(\hat p,\hat q)$.}
Substituting the values \eq{ZZ_neutral1}--\eq{ZZ_neutral3} 
into $\Gamma_{ab}$, we obtain 
the partition functions of neutral ZZ branes: 
\begin{align}
 \Gamma^{(11)}_{ab}(\tau^*)
  =-(-1)^{m+n}\,\frac{2\,\beta\,\hat q\,\hat p}{\hat q^2-\hat p^2}
 \,a^{\hat q+\hat p}\,\sin\frac{m(\hat q-\hat p)}{2\hat p}\pi\cdot
 \sin\frac{n(\hat q-\hat p)}{2\hat q}\pi. 
\end{align}
They completely agree with the analysis of \cite{SeSh}.

\subsection{Charged ZZ branes}

The $\Gamma^{(12)}_{ab}$ is expressed 
as a function of $\tau$ 
by integrating $d\Gamma^{(12)}_{ab}/d\tau
=(d\Gamma^{(12)}_{ab}/d\zeta)\cdot(d\zeta/d\tau)$. 
Noting that
\begin{align}
 \frac{d\Gamma^{(12)}_{ab}}{d\zeta}
  =Q^{(1)}_a\bigl(\zeta(\tau)\bigr)+Q^{(1)}_b\bigl(-\zeta(\tau)\bigr) 
  =Q^{(1)}_0\bigl(\zeta(\tau_a)\bigr)
   +Q^{(1)}_0\bigl(\zeta(\bar\tau_b)\bigr)
 \label{saddle_pt_eq_charged}
\end{align} 
with $\tau_a\equiv \tau+2\pi ia/\hat p$ 
and $\bar\tau_b\equiv \tau+\pi i\, (2b+1)/\hat p$,
we obtain
\begin{align}
 \Gamma_{ab}^{(12)}
 &=\beta\,\hat p\,a^{\hat p+\hat q}
  \biggl[\, 
  \frac{1}{\hat q+\hat p}\Bigl(\cosh(\hat p\tau+\hat q\tau_a\bigr)
   +\cosh\bigl(\hat p \tau +\hat q \bar\tau_b\bigr)\Bigr) \mp \nn\\
 &~~~~~~~~~~~~\mp \frac{1}{\hat q-\hat p}\Bigl(\cosh(\hat p\tau-\hat q\tau_a\bigr)
   +\cosh\bigl(\hat p \tau-\hat q\bar\tau_b\bigr)\Bigr)\biggr],
\end{align}
where the upper ({\em resp}.\ lower) sign corresponds
to 1-cut ({\em resp}.\ 2-cut) solutions.  
The saddle points are found similarly by solving \eq{saddle_pt_eq_charged} 
as 
\begin{align}
 \tau^*=\begin{cases} 
  \dfrac{1}{2}\,\Bigl(-\dfrac{2(a+b)-1}{\hat p}+\dfrac{2k+1}{\hat q}\Bigr)\,
   \pi i & (\mbox{1-cut}) \\
  \dfrac{1}{2}\,\Bigl(-\dfrac{2(a+b)-1}{\hat p}+\dfrac{2k}{\hat q}\Bigr)\,
   \pi i \quad & (\mbox{2-cut})
 \end{cases}
 \quad (k\in\bZ).
 \label{ZZ_charged1}
\end{align}
Setting $m=2(b-a)+1$ for both of 1-cut and 2-cut solutions, 
and $n=2k+1$ for 1-cut and $n=2k$ for 2-cut solutions, 
we have
\begin{align}
 \tau^*_a&=\frac{1}{2}\Bigl(-\frac{m}{\hat p}+\frac{n}{\hat q}\Bigr)\,\pi i,
 \label{ZZ_charged2}\\
 \bar\tau^*_b&=\frac{1}{2}\Bigl(+\frac{m}{\hat p}+\frac{n}{\hat q}\Bigr)\,\pi i.
 \label{ZZ_charged3}
\end{align}
The number of these charged ZZ branes can be counted in a similar way, 
and we obtain 
\begin{align}
\begin{cases} 
  \Bigl[\dfrac{\hat p\hat q+1}{2}\Bigr]\times 2 & (\mbox{1-cut}) \\
  \Bigl[\dfrac{\hat p(\hat q-1)+1}{2}\Bigr]\times 2  & (\mbox{2-cut}).
 \end{cases}
\end{align}
Substituting the values \eq{ZZ_charged1}--\eq{ZZ_charged3} 
into $\Gamma_{ab}$, we obtain 
the partition functions of charged ZZ branes: 
\begin{align}
 \Gamma^{(12)}_{ab}(\tau^*)
  =-(-1)^{m+n}\,\frac{2\,\beta\,\hat q\,\hat p}{\hat q^2-\hat p^2}
 \,a^{\hat q+\hat p}\,\sin\frac{m(\hat q-\hat p)}{2\hat p}\pi\cdot
 \sin\frac{n(\hat q-\hat p)}{2\hat q}\pi. 
\end{align}
They also agree with the analysis of \cite{SeSh}.

\section{Conclusion and discussion}

In this paper, we formulate a string field theory 
of type 0B $(p,q)$ minimal superstrings, 
based on the Douglas equation of two-cut two-matrix models. 
We find that the 2cKP hierarchy is the underlying integrable 
structure in these systems, 
and solutions are given by decomposable fermion states 
which satisfy the $W_{1+\infty}$ constraints. 
We show that the 2cKP hierarchy is reducible 
for backgrounds invariant under the $\bZ_2$ transformation, 
$\tilde{\rm C}:(\bsigma,\bL)\to(-\bsigma,-\bL)$, 
which is the case for odd minimal superstrings. 
Taking into account this reduction, 
we establish the correspondence between the operators in the 2cKP hierarchy 
and those of super Liouville theory. 
We also show that background R-R fluxes are naturally 
incorporated in our treatment without ambiguity. 
We calculate various correlation functions 
and show that they all agree with the results 
of super Liouville theory.

Type 0A superstrings are obtained by orbifolding 
the above type 0B systems with another $\bZ_2$ symmetry 
$C:(\bsigma,\bL)\to(-\bsigma,\bL)$ (charge conjugation). 
This can be carried out straightforwardly in the absence of R-R flux. 
However, in the presence of R-R fluxes, 
we better move to the description using Toda lattice hierarchy. 
The investigation in this direction is in progress, 
and will be reported in our future communication \cite{fi2}.

\section*{Acknowledgments}

We thank Yoshinori Matsuo, Shigenori Seki and 
Takao Suyama for useful discussions. 
This work was supported in part by the Grant-in-Aid for 
the 21st Century COE
``Center for Diversity and Universality in Physics'' 
from the Ministry of Education, Culture, Sports, Science 
and Technology (MEXT) of Japan. 
MF and HI are also supported by the Grant-in-Aid for 
Scientific Research No.\ 15540269 and No.\ 18\textperiodcentered 2672, 
respectively, from MEXT.

\appendix
\section{Derivation of eq.\ (\ref{PandQ})}\label{Douglas}

To study the double scaling limit, 
we interpret the operators $\bQone,\,\bQtwo,...$ as 
the following difference operators with respect to the index $n=0,1,\cdots$:
\begin{align}
 \bra{\alpha_n} \bQone^{\rm T}=\sum_{m}\bigl(\bQone^{\rm T}\bigr)_{nm}
  \bra{\alpha_m}\,
  =\sum_{k=m-n}\,\bigl(\bQone^{\rm T}\bigr)_{n,n+k}
  \,e^{k\partial_n}\,\bra{\alpha_n}
  \equiv \sum_{k} \, a_{k}(n)\,\Lambda^{k}\bra{\alpha_n}, 
\end{align} 
where $a_{k}(n)\equiv (\bA^{\rm T})_{n,n+k}$ 
and $\Lambda\,f(n)=f(n+1)$. 
Equation \eq{douglas_lat} (or \eq{EOMofPQ}) 
then gives the recursion relations for the sequence 
$\bigl\{a_k(n)\bigr\}_{n=0}^\infty \ (k\in \mathbb Z)$.

In the vicinity of the point $(x_c,y_c)=(0,0)$ we consider, 
the orthonormal polynomials 
$\bigl\{ \alpha_n(x),\, \beta_n(y)\bigr\}_{n=0,1,2,\cdots}$ 
do not behave smoothly for a small change of $n$, 
and thus we need to redefine them as follows \cite{multcmultc}:
\begin{align}
\begin{pmatrix}
\widetilde{\alpha}_{2k}^{(e)}(x) \\
\widetilde{\alpha}_{2k+1}^{(o)}(x) 
\end{pmatrix}
\equiv 
(-1)^k
\begin{pmatrix}
\alpha_{2k}(x) \\
\alpha_{2k+1}(x)
\end{pmatrix}
,\quad 
\begin{pmatrix}
\widetilde{\beta}_{2k}^{(e)}(x) \\
\widetilde{\beta}_{2k+1}^{(o)}(x)
\end{pmatrix}
\equiv 
(-1)^k\,
\begin{pmatrix}
\beta_{2k}(x) \\ 
\beta_{2k+1}(x) 
\end{pmatrix}.
\end{align}
With this new basis, a smooth operator acting on orthonormal polynomials 
can be expressed as a $2\times 2$ matrix of the following form:
\begin{align}
\widetilde{\bQone}\equiv 
(-1)^{[\frac{\boldsymbol{n}}{2}]}\bQone (-1)^{[\frac{\boldsymbol{n}}{2}]}
=\sum_{l=0}^{\infty}(-1)^l
\left(
\begin{matrix}
a_{-2l}(2k)\, \Lambda^{-2l} & a_{-2l+1}(2k+1)\,\Lambda^{-2l+1} \\
-a_{-2l+1}(2k)\,\Lambda^{-2l+1} & a_{-2l}(2k+1)\, \Lambda^{-2l}
\end{matrix}
\right),
\end{align}
where $\boldsymbol{n}=(n\,\delta_{mn})$, 
and the index is rearranged 
to run as $(0,2,4,\cdots;1,3,5,\cdots)$.

For a generic case, the even and odd subsequences 
($\bigl\{a_l(2k)\bigr\}_{k=0}^\infty$ and 
$\bigl\{a_l(2k+1)\bigr\}_{k=0}^\infty$) 
are allowed to have different large $N$ limits.
At the 2-cut critical points, however, since the system carries 
a continuum phase transition to 1-cut phase, 
we assume that both of the even/odd subsequences have the same 
limit $a_l^c$.
Thus the leading large $N$ behaviour of $\widetilde{\bQone}$ 
can be given as 
\begin{align}
 \widetilde{\bQone}\sim \sum_{l=0}^\infty (-1)^l
  \Big(a_{-2l}^c\,\Lambda^{-2l} 
  -i\sigma_2\,a_{-2l+1}^c\,\Lambda^{-2l+1}\Big).
\end{align}
If we assume that this behaves as $\sim(\Lambda-1)^{\hat p}$ 
with some integer $\hat{p}\in \mathbb N$, 
$\widetilde{\bQone}$ becomes a differential operator 
of order $\hat{p}$. 
The same also holds for $\widetilde{\bQtwo}$, 
which we assume to give a differential operator of order $\hat q$. 
The correctness of this assumption can be checked 
by using eq.\ \eq{EOMofPQ} 
with the corresponding critical potentials $V_1(x)$ and $V_2(y)$, 
and can actually be shown 
by solving the linear equations 
for the coefficients of potentials 
as in the case of bosonic strings \cite{2-mat}.

\section{Charge conjugation in terms of infinite dimensional Grassmannian}
\label{C_Grassmannian}

The charge conjugation  is realized by exchanging the operators 
$\psi^{(1)}(\lambda)$ and $\psi^{(2)}(\lambda)$. 
This transformation is well described 
in terms of the Sato-Wilson operator $\bW(x;\del)$. 
It is defined by the similarity transformation with $\sigma_1$ as 
\begin{align}
\bW(x;\del)\to \bW^{\rm C}(x^{\rm C};\del)\equiv \sigma_1\bW(x;\del)\sigma_1.
\end{align}
Under this transformation, the Sato equations are invariant,
\begin{align}
g\,\frac{\del \bW^{\rm C}(x^{\rm C};\del)}{\del x^{\rm C}_n}
= -\bigl(\bW^{\rm C}\sigma_3^\mu\del^n\bW^{\rm C}{}^{-1}\bigr)_-\bW^{\rm C}(x^{\rm C};\del).
\end{align}
The pair of differential operators $(\bP,\bQ)$ at the background $b$ 
transforms as
\begin{align}
\bigl(\bP(b),\bQ(b)\bigr) 
\to 
\bigl(-\bP^{\rm C}(b^{\rm C}),-\bQ^{\rm C}(b^{\rm C})\bigr),
\end{align}
and eq.\  \eq{wave-baker} is changed into the following: 
\begin{align}
\bP^{\rm C}\,\tilde\Psi^{\rm C}(x^{\rm C};\zeta)
=\tilde\Psi^{\rm C}(x^{\rm C};\zeta)\, 
 \begin{pmatrix}
     -\zeta & 0 \cr 0 & \zeta
    \end{pmatrix},\quad 
\bQ^{\rm C}\,\tilde\Psi^{\rm C}(x^{\rm C};\zeta)
=g\,\tilde\Psi^{\rm C}(x^{\rm C};\zeta)
\,\begin{pmatrix}
     -\,\overleftarrow{\frac{\del}{\del\zeta}} & 0 
     \cr 0 & \overleftarrow{\frac{\del}{\del\zeta}}
    \end{pmatrix}
\label{wavefuncC}
\end{align}
with 
$\Psi^{\rm C}(x^{\rm C};\zeta)
\equiv\sigma_1\Psi(x^{\rm C};\zeta)\sigma_1$.

We now turn to the free-fermion description. 
The charge conjugation of a decomposable fermion state $\ket{\Phi}$ 
is given by%
\footnote{
Recall that $\bar\psi(\lambda)$ is a column vector, 
$\bar\psi(\lambda)=\bigl( \bar\psi^{(1)}(\lambda),\bar\psi^{(2)}(\lambda)
\bigr)^{\rm T}$.
} 
\begin{align}
\ket{\Phi}&=\prod_{k=0}^\infty\prod_{i=1}^2
\Bigl[
\oint \frac{d\lambda}{2\pi i}\,
\bPhi_k^{(i)}(x=0;\lambda)\,
\bar\psi(\lambda)
\Bigr]
\ket{\infty} \nn\\
\to&
\prod_{k=0}^\infty\prod_{i=1}^2
\Bigl[
\oint \frac{d\lambda}{2\pi i}\,
\bPhi_k^{{\rm C}(i)}(x^{\rm C}=0;\lambda)\,
\bar\psi(\lambda)
\Bigr]
\ket{\infty}
=\prod_{k=0}^\infty\prod_{i=1}^2
\Bigl[
\oint \frac{d\lambda}{2\pi i}\,
\bPhi_k^{(i)}(x=0;\lambda)\,\sigma_1
\bar\psi(\lambda)
\Bigr]
\ket{\infty} \nn\\
\equiv& \ket{\Phi^{\rm C}},
\end{align}
and we see that the two fermions, $\psi^{(1)}(\lambda)$ and 
$\psi^{(2)}(\lambda)$, are actually exchanged.

The charge conjugation is not the symmetry of the theory 
and changes the background of the theory in general. 
If we consider, however, the $\mathbb Z_2$ symmetric matrix models, 
$w(-x,-y)=w(x,y)$, this system is $\mathbb Z_2$ symmetric at least 
when their two Fermi levels are equal to each other. 
So we can require the following $\mathbb Z_2$ symmetric property:
\begin{align}
\ket{\Phi^{\rm C}}=\ket{\Phi}
\end{align}
at least perturbatively.

\section{Derivation of string equations}\label{string_eqs}
\label{string_eqn}

In this appendix we derive the string equations 
\eq{se_p1}--\eq{se_p2b} from the Virasoro constraints \eq{string_eq0}. 
First we introduce a linear differential operator $\cO$ as
\begin{align}
 \cO\equiv \sum_{n\ge 1}
 (n+\hat p)
 \bigl(x_{n+\hat p}^{(1)}\del_n^{(1)}-x_{n+\hat p}^{(2)}\del_n^{(2)}\bigr),
 \label{lntau}
\end{align}
then \eq{string_eq0} can be rewritten as  
\begin{align}
\cO \ln \tau_\nu (x) 
&=-\frac{1}{2g^2}\sum_{n=1}^{\hat p-1}
 n(\hat p-n)
 \bigl(x_{\hat p-n}^{(1)}x_n^{(1)}-x_{\hat p-n}^{(2)}x_n^{(2)}\bigr)
 -\frac{\hat p \nu}{g} \bigl(x_{\hat p}^{(1)}+x_{\hat p}^{(2)}\bigr),
\end{align}
which takes for $\hat p=1,\,2$ as
\begin{align}
 \hat p=1\ :&\quad \cO \ln \tau_\nu(x)
  =-\frac{\nu}{g}\, \bigl(x_1^{(1)}+x_1^{(2)}\bigr) 
 \label{C2},
 \\
 \hat p=2\ :&\quad \cO \ln \tau_\nu(x)=
 -\frac{1}{2g^2}\Bigl((x_1^{(1)})^2-(x_1^{(2)})^2\Bigr)
 -\frac{2\nu}{g}\,\bigl(x_2^{(1)}+x_2^{(2)}\bigr).
 \label{C3}
 \end{align}
Furthermore, taking a difference between eq.\ \eq{lntau} with $\nu\pm1$ 
and that with $\nu$, 
we obtain the following formula for $H_\pm(x)=\tau_{\nu\pm1}(x)/\tau_\nu(x)$:
\begin{align}
 \cO\,\ln H_\pm(x)=\mp\,\frac{\hat p}{g}\,(x^{(1)}_{\hat p}+x^{(2)}_{\hat p}). 
\end{align}
Then we obtain string equations as follows. 
First for the $\hat p=1$ case, 
by using eqs.\ \eq{rel1a}, \eq{rel1b} and \eq{rel2} 
we obtain  
\begin{align}
&\sum_{n\ge 1}(n+1)\Bigl(x_{n+1}^{(1)}(\be^{(1)}\bL^n)_{-1}
-x_{n+1}^{(2)}(\be^{(2)}\bL^n)_{-1}\Bigr)
=-g\,\cO w_1 \nn\\
& =-g\,\cO \cdot 
\begin{pmatrix}
-g\,\del_1^{(1)}\ln \tau_\nu & -H_+ \cr
H_- & -g\,\del_1^{(2)}\ln\tau_\nu
\end{pmatrix} 
=
\begin{pmatrix}
g^2\,\del_1^{(1)}\cO\ln \tau_\nu 
& 
g\,H_+\,\cO \ln H_+
\cr
-g\,H_-\,\cO \ln H_- 
&
g^2\,\del_1^{(2)}\cO\ln\tau_\nu
\end{pmatrix}\nn\\
&=
\begin{pmatrix}
-g\,\nu  &  -H_+(x_1^{(1)}+x_1^{(2)})  \cr
-H_-(x_1^{(1)}+x_1^{(2)})  &  -g\,\nu
\end{pmatrix} =-g\,\nu\bunit_2
-\Bigl(x_1^{(1)}(\be^{(1)})_{-1}-x_1^{(2)}(\be^{(2)})_{-1}\Bigr),
\end{align}
from which we obtain the string equation \eq{se_p1}. 
In deriving the last expression, we used the formula \eq{form_e1}.

As for the $\hat p=2$ case, 
by using eq.\ \eq{rel1a} and \eq{rel2} again we have 
\begin{align}
&\!\!\!\!\!\!\!\sum_{n\ge 1}(n+2)\bigl(x_{n+2}^{(1)}(\be^{(1)}\bL^n)_{-1}
-x_{n+2}^{(2)}(\be^{(2)}\bL^n)_{-1}\bigr) \nn\\
&=
\begin{pmatrix}
-x_1^{(1)}  &  -2H_+(x_1^{(1)}+x_1^{(2)})  \cr
-2H_-(x_1^{(1)}+x_1^{(2)})  &  x_1^{(2)}
\end{pmatrix} \nn\\
&=-2\Bigl(x_2^{(1)}(\be^{(1)})_{-1}-x_2^{(2)}(\be^{(2)})_{-1}\Bigr)
-\Bigl(x_1^{(1)}(\be^{(1)}\bL^{-1})_{-1} -x_1^{(2)}(\be^{(2)}\bL^{-1})_{-1}\Bigr),
\end{align}
from which we obtain the first string equation \eq{se_p2a}. 
The second string equation \eq{se_p2b} 
is obtained with a little algebra, but in a way similar to the above:  
\begin{align}
 &\!\!\!\sum_{n\ge 1}(n+2)\bigl(x_{n+2}^{(1)}(\be^{(1)}\bL^n)_{-2}
 -x_{n+2}^{(2)}(\be^{(2)}\bL^n)_{-2}\bigr)  
 =-g\,\cO w_1+g\,\cO w_1\cdot w_1
 \nn\\
&=
 -g
 \begin{pmatrix}
 \nu -1/2 & 0 \cr 0 & \nu+1/2
 \end{pmatrix}\nn\\
&~~~~
 +2(x_2^{(1)}+x_2^{(2)})
 \begin{pmatrix}
  H_+ H_- & g\,\del^{(2)}_1 H_+ \cr
  g\,\del^{(1)}_1 H_- & -H_+ H_-
 \end{pmatrix}
 -(x_1^{(1)}+x_1^{(2)})
 \begin{pmatrix}
  0   & H_+ \cr
  H_- & 0
 \end{pmatrix}
\nn\\
&=
 -g
 \begin{pmatrix}
 \nu -1/2 & 0 \cr 0 & \nu+1/2
 \end{pmatrix}\nn\\
&~~~~ 
-2\Bigl[x_2^{(1)}(\be^{(1)})_{-2}-x_2^{(2)}(\be^{(2)})_{-2}\Bigr] 
-\Bigl[x_1^{(1)}(\be^{(1)}\bL^{-1})_{-2} 
 -x_1^{(2)}(\be^{(2)}\bL^{-1})_{-2}\Bigr],
\end{align}
where we have used the formulas \eq{form_e2} and \eq{form_eL2}.

\section{ZS hierarchy and $\hat p=1$ superstrings }\label{ZS}

Critical behaviour of 2-cut one matrix models have been studied extensively, 
and a series of string equations are obtained explicitly 
\cite{GroWit,PeShe,Napp,CDM,HMPN,ogu,BDJT,UniCom,SeSh2}. 
We here rederive the results 
from the viewpoint of 2cKP hierarchy, 
and compare them with our analysis based on the string field formulation.

We first note that any 2-cut one-matrix model can be 
realized as a 2-cut two-matrix model 
by taking $V_2(y)=cy^2/2$ in the potential 
$w(x,y)=V_1(x)+V_2(y)-cxy$. 
Then the differential operator $\bP$ is of order one 
and will have the form
\begin{align}
\bP=\sigma_3\,\del+2H,\qquad H\equiv
\begin{pmatrix}
0 & H_+ \cr H_- & 0
\end{pmatrix},
\end{align}
where $H_\pm\in i\,\bR$ due to the reality condition \eq{OpHer}. 
This corresponds to the so-called Zakharov-Shabat hierarchy \cite{CDM,HMPN} 
and was proposed to describe $\hat p=1$ minimal superstrings 
in \cite{UniCom}. 
The reduction condition 
$\bP=\bsigma\bL=(\bsigma\bL)_+$ is in our language 
the $\hat p=1^{\rm st}$ reduction 
of 2cKP hierarchy 
with $H_\pm=\tau_{\nu\pm1}/\tau_\nu$.  
The pseudo-differential operator 
$\bsigma=\sigma_3+2\sum_{n=1}^\infty H_n\,\del^{-n}$ 
can be easily obtained 
by solving the conditions 
$[\bP,\bsigma]=0$ and $\bsigma^2=\bunit_2$ as 
\begin{align}
H_1&=H,\qquad H_2=-\frac{1}{2}\del H-\sigma_3H^2,\nn\\
H_3&=\frac{3}{2}\sigma_3H\del H+\frac{1}{2}\sigma_3\del H\cdot H
+\frac{1}{4}\del^2H-2H^3, \label{H_n}\\
H_4&=-\frac{7}{4}\sigma_3H\del^2H-\frac{5}{4}\sigma_3(\del H)^2
-\frac{1}{4}\sigma_3\del^2H\cdot H+ 3\sigma_3H^4+ \nn\\
 &\quad +6H^2\del H+3H\del H\cdot H-\frac{1}{8}\del^3H,\ \cdots.\nn
\end{align}
From  this one finds
\begin{align}
 (\bsigma)_{-1}&=2H, \\
 (\bsigma\bP)_{-1}&= 2H^2-\del H,\\
 (\bsigma\bP^2)_{-1}&=\sigma_3\cdot( \del H\cdot H-H\cdot\del H)
  +\frac{1}{2}\,\del^2 H+4H^3,\\
 (\bsigma\bP^3)_{-1}
  &=\frac{1}{2}\,\del^2 H\cdot H+6 H^4
    +\frac{1}{2}\,H\cdot\del^2 H-\frac{1}{2}\,(\del H)^2
    +\sigma_3\,\Bigl(\frac{1}{4}\,\del^3 H+6H^2\cdot\del H\Bigr).
\end{align}
The differential operator $\bQ$ is then given by
\begin{align}
 \bQ=\sum_{n=1}^{\hat q+1}n\,b_n^{[n-1]}\,(\bsigma\bP^{n-1})_+
  =b^{[0]}_1\sigma_3+\sum_{n=2}^{\hat q+1}n\,b_n^{[n-1]}
   \,(\bsigma\bP^{n-1})_+,
\end{align}
and the corresponding string equation is obtained 
by rewriting the Douglas equation as%
\footnote{
Setting $b^{[0]}_1=\xi$, we have $[\sigma_3\del, \sigma_3 b^{[0]}_1]
 =g\bunit_2$, 
and thus the Douglas equation $[\bP,\bQ]=g\bunit_2$ 
reduces to \eq{StrEq-p1}
} 
\begin{align}
 0&=\sum_{n=1}^{\hat q+1}n\,b_n^{[n-1]}\bigl[ \bP,(\bsigma\bP^{n-1})_+ \bigr] 
 =-\sum_{n=1}^{\hat q+1}n\,b_n^{[n-1]}
  \bigl[ \bP,(\bsigma\bP^{n-1})_- \bigr] \nn\\
 &=-\sum_{n=1}^{\hat q+1}n\,b_n^{[n-1]}
  \bigl[ \sigma_3\del + 2H,
  (\bsigma\bP^{n-1})_{-1}\del^{-1}+(\bsigma\bP^{n-1})_{-2}\del^{-2}
   +\cdots \bigr] \nn\\
 &=-\sum_{n=1}^{\hat q+1}n\,b_n^{[n-1]}
  \bigl[ \sigma_3,(\bsigma\bP^{n-1})_{-1} \bigr].
\label{StrEq-p1}
\end{align}
Since a nontrivial physical operator appears only once for each $n$, 
we use the parametrization $t_n\equiv b_n^{[n-1]}$ as in the main text.

A string equation of the $\hat q=2$ case is given by
\begin{align}
\left\{
\begin{array}{ll}
0=-2t_1H_+-3t_3 \bigl(\frac{1}{2}\del^2H_++4H_+^2H_-\bigr), \\
0=-2t_1H_--3t_3 \bigl(\frac{1}{2}\del^2H_-+4H_-^2H_+\bigr).
\end{array}
\right.
\label{str_App}
\end{align}
By introducing $H_\pm=\frac{i}{4}(f_1\pm f_2)$ and  
taking $t_3\equiv 8/3$ $(\beta=4)$ and $t_1\equiv -\mu$, 
eqs.\ \eq{str_App} are rewritten as 
\begin{align}
\left\{
\begin{array}{ll}
0=\frac{1}{2}\mu f_1-g^2f_1''+\frac{1}{2}f_1(f_1^2-f_2^2), \\
0=\frac{1}{2}\mu f_2-g^2f_2''+\frac{1}{2}f_2(f_1^2-f_2^2).
\end{array}
\right.
\end{align}
Here $'$ implies the derivative with respect to $\mu$, 
and we have used $\del=g\,\del/\del t_1=-g\,\del/\del\mu$. 
It is also convenient to rewrite them 
with $H_\pm\equiv \frac{i}{4}r e^{\pm \chi}$ as 
\begin{align}
\left\{
\begin{array}{ll}
0=\dfrac{1}{2}\mu r-g^2 r''-g^2 r(\chi')^2+\dfrac{1}{2}r^3, 
\\
0=(r^2\chi')'.
\end{array}
\right.
\label{str_eq_12a}
\end{align}
Noticing that $r^2\chi'$ can be written 
as $-(1/g)\,(H_+\del H_- -\del H_+ H_-)$, 
one can see that the first equation of \eq{str_eq_12}
gives an integration constant for the second equation of 
\eq{str_eq_12a} as $r^2\chi'=\nu$. 
Taking the $\mathbb Z_2$ symmetric reduction 
($\chi=0$ and thus $\nu=0$) gives 
the celebrated Painlev\'e II equation \cite{PeShe}:
\begin{align}
0=\frac{1}{2}\mu r-g^2 r''+\frac{1}{2}r^3.
\end{align}

The $\hat q=3$ critical point exists in a flow 
generated by an R-R operator \cite{UniCom}. 
The string equation is given by 
\begin{align}
\left\{
\begin{array}{ll}
0=-2t_1H_+-4t_4 \bigl(\frac{1}{4}\del^3H_++6H_+H_-\del H_+\bigr), \\
0=-2t_1H_-+4t_4 \bigl(\frac{1}{4}\del^3H_-+6H_+H_-\del H_-\bigr),
\end{array}
\right.
\end{align}
and is written with 
$H_\pm=\frac{i}{4}(f_1\pm f_2)$, $t_4\equiv 4$ and $t_1\equiv -\mu$
as
\begin{align}
\left\{
\begin{array}{ll}
0=\frac{1}{2}\mu f_1-f_2'''+\frac{3}{2}f_2'(f_1^2-f_2^2), \\
0=\frac{1}{2}\mu f_2-f_1'''+\frac{3}{2}f_1'(f_1^2-f_2^2).
\end{array}
\right.
\end{align}
This is the string equation found in \cite{CDM,HMPN}. 
We can also rewrite it as 
\begin{align}
\left\{
\begin{array}{rl}
0\!\!\!&=\frac{1}{2}\mu r-3(r'\chi')'
-r(\chi'''+(\chi')^3)+\frac{3}{2}r^3\chi', \\
0\!\!\!&=\bigl(rr''-\frac{1}{2}(r')^2+\frac{3}{2}r^2(\chi')^2
-\frac{3}{13}r^4\bigr)'.
\end{array}
\right.
\end{align}
The R-R flux $\nu$ again gives its integration constant.

We end this Appendix with a comment 
on the normalization of the free energy $F(\mu;g)$. 
By using the formula \eq{toda} for $H_\pm$ 
together with the reduction condition $\del^{[1]}_1\tau_\nu=0$, 
we obtain 
\begin{align}
 g^2(\del_1^{[0]})^2\ln \tau_\nu
 =g^2\frac{\del^2}{\del \mu^2}\ln\tau_\nu=4H_+H_-
 =-\frac{1}{4}\,r^2. 
 \label{Hirotap1}
\end{align}
It is consistent with the normalization of one-matrix-model calculation 
\cite{PeShe,Napp,CDM,HMPN}, 
\begin{align}
u(\mu;g)\equiv \frac{\del^2}{\del \mu^2}F(\mu;g)
 =\frac{1}{4}r^2(\mu;g),
\end{align}
with the identification $\cZ=e^{-g^{-2}F(\mu;g)}=\tau_\nu(\mu;g)$ \cite{HMPN}.

\setlength{\itemsep}{5.\baselineskip}

\end{document}